\documentclass[aps,prx,twocolumn,superscriptaddress, longbibliography]{revtex4-2}
\usepackage{amsmath}    \usepackage{graphicx}
\UseRawInputEncoding
\usepackage{booktabs}
\usepackage{xcolor,soul}
\usepackage{amssymb}
\usepackage{mathdots}
\definecolor{darkblue}{HTML}{3771C8}
\usepackage[colorlinks=true,
    pdfborder={0 0 0},
    linkcolor=darkblue,
    citecolor = darkblue, 
    urlcolor = darkblue
]{hyperref}
\usepackage[capitalize,nameinlink]{cleveref}
\usepackage[normalem]{ulem}
\usepackage{braket}

\newcommand{\exc}[1]{\langle #1 \rangle}

\renewcommand{\selectlanguage}[1]{}
\begin{document}
\title{Stabilizing remote entanglement via waveguide dissipation}

\author{Parth S Shah}
    \email[These authors contributed equally.]{}
    \affiliation{Moore Laboratory of Engineering, California Institute of Technology, Pasadena, California 91125}
    \affiliation{Institute for Quantum Information and Matter, California Institute of Technology, Pasadena, California 91125}

\author{Frank Yang}
    \email[These authors contributed equally.]{}
    \affiliation{Moore Laboratory of Engineering, California Institute of Technology, Pasadena, California 91125}
    \affiliation{Institute for Quantum Information and Matter, California Institute of Technology, Pasadena, California 91125}

\author{Chaitali Joshi}
    \email[These authors contributed equally.]{}
    \affiliation{Moore Laboratory of Engineering, California Institute of Technology, Pasadena, California 91125}
    \affiliation{Institute for Quantum Information and Matter, California Institute of Technology, Pasadena, California 91125}

\author{Mohammad Mirhosseini}
    \email[mohmir@caltech.com; http://qubit.caltech.edu]{}
    \affiliation{Moore Laboratory of Engineering, California Institute of Technology, Pasadena, California 91125}
    \affiliation{Institute for Quantum Information and Matter, California Institute of Technology, Pasadena, California 91125}

\date{\today} 

\begin{abstract} Distributing entanglement between remote sites is integral to quantum networks. Here, we demonstrate the autonomous stabilization of remote entanglement between a pair of non-interacting superconducting 
qubits connected by an open waveguide on a chip. In this setting, the interplay between a classical continuous drive - supplied through the waveguide - and dissipation into the waveguide stabilizes the qubit pair in a dark state, which, asymptotically, takes the form of a Bell state. We use field-quadrature measurements of the photons emitted to the waveguide to perform quantum state tomography on the stabilized states, where we find a concurrence of $0.504^{+0.007}_{-0.029}$ in the optimal setting with a stabilization time constant of 56 $\pm$ 4 ns. We examine the imperfections within our system and discuss avenues for enhancing fidelities and achieving scalability in future work. The decoherence-protected, steady-state remote entanglement offered via dissipative stabilization may find applications in distributed quantum computing, sensing, and communication.

\end{abstract}
\maketitle
\subsection*{Introduction} Entanglement between distant physical systems is a crucial resource for quantum information processing. Over long distances, entanglement can make communication secure against eavesdropping and resilient to loss \cite{Briegel:1998jd,10.1103/physrevlett.67.661,Wehner:2018cu}. On shorter length scales, entanglement between distant non-interacting modules
can help realize non-local gate operations in a quantum computer \cite{10.1038/46503,10.1103/physreva.76.062323,10.1103/physreva.62.052317,Monroe:2014fz}. Remote entanglement can be created by the deterministic exchange of photons between qubits located at remote sites (e.g., \cite{10.1038/s41586-023-05885-0,10.1038/s41567-019-0507-7}). Alternatively, quantum measurements can be used to `herald' entanglement in a probabilistic fashion (e.g., \cite{Duan:2001dt,Roch:2014ey}). Once established, entangled states have to be protected from decoherence by the environment before they can be used, a task that can be achieved via passive storage in isolated quantum memories \cite{Wehner:2018cu}. Current research actively pursues the development of physical systems capable of high-bandwidth generation, distribution, and storage of remote entanglement. 

An entirely different approach for the generation and protection of entanglement is its \emph{stabilization}. This method employs dissipation into a shared reservoir, in combination with continuous drives, to establish entanglement between two or more parties. Intriguingly, such an engineered dissipation can not only create entanglement but also protect it - indefinitely - from environment-induced errors \cite{10.1103/physrevlett.77.4728,Verstraete:2009kc}. `Driven-dissipative' processes thus provide an attractive route for the generation and autonomous preservation of entanglement \cite{10.1103/physreva.59.2468,10.1103/physrevlett.107.080503,10.1103/physrevlett.106.090502,10.1038/nature12801,10.1038/nature12802,10.1103/physrevlett.117.040501,10.1103/physrevlett.116.240503,10.1103/physrevx.6.011022,10.1038/s41467-022-31638-0}. Beyond entanglement generation, dissipative processes have also been studied for a variety of other tasks in quantum information processing as an alternative to unitary gate operations \cite{10.1038/nphys1073,Verstraete:2009kc,10.1103/physrevlett.115.200502,10.1038/nature09801}. Despite the wide interest in this area, however, stabilizing \emph{remote} entanglement has remained elusive, primarily due to the challenge of engineering shared dissipation for remote sites.

Spontaneous emission into a one-dimensional photonic bath can provide a shared dissipation channel for remote quantum emitters. Such a system can be realized within the paradigm of waveguide quantum electrodynamics (QED), where two or multiple qubits - acting as quantum emitters - are strongly coupled to a shared waveguide. In this setting, the interference of photons emitted by the qubits can give rise to the formation of collective \emph{dark states} that are protected from dissipation by their internal symmetries \cite{vanLoo2013Dec,Mirhosseini2019May, zanner2022, sheremet2021,10.1126/science.ade9324}. Theoretical work has proposed a variety of methods for stabilizing collective dark states \cite{10.1103/physreva.41.3782,PhysRevLett.92.013602,10.1103/physrevlett.110.040503,10.1088/1367-2630/14/6/063014,10.1103/physrevlett.113.237203,10.1038/s41534-017-0020-8,10.1103/physreva.97.023810,10.1103/physreva.98.012329,10.1103/physrevresearch.4.023010,10.1103/physreva.105.062454, Rabl2023,10.1103/physreva.98.012329,10.1103/physrevresearch.5.013127}. However, experimental demonstrations of these proposals have remained out of reach, owing to the need for components like the injection of non-classical states into the waveguide and the requirements for either unidirectional or time-modulated qubit-photon coupling -- elements that present practical challenges to implement.

\begin{figure*}[t!]
\centering
\includegraphics[width=\linewidth]{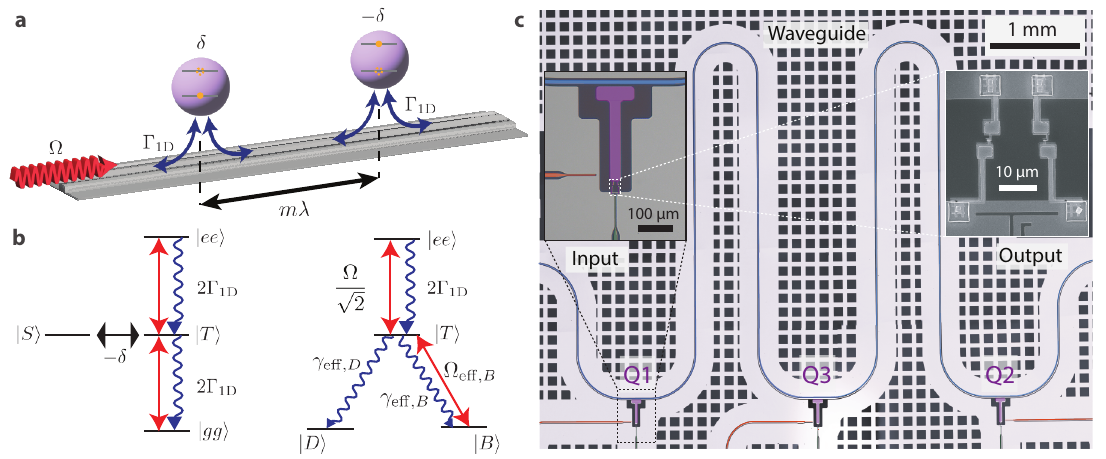}
\caption{\textbf{Experimental setup.} a) A pair of quantum emitters coupled to a shared waveguide. The qubit transition frequencies are offset in opposite directions with respect to a central frequency, and their separation is set equal to the wavelength of the radiation at their center frequency. A continuous Rabi drive is supplied through the channel. b) Left: The energy level diagram of the system. The triplet state $|T\rangle$ superradiantly decays into the waveguide. The singlet state $|S\rangle$ is coherently coupled to the triplet but has no direct decay path into the waveguide. Right: The energy diagram in a basis including the dark state $|D\rangle$. Here, the population is pumped from the triplet state into the dark state, which is protected from decay into the waveguide. c) Optical image of the fabricated device, where three transmon qubits are coupled to a shared coplanar waveguide on a chip. We only utilize qubits 1 and 2 in this experiment. Each qubit can be addressed via flux and charge drives delivered via a pair of on-chip control lines. }
\label{figure 1}
\end{figure*}

Here, we demonstrate the \emph{stabilization of remote entanglement} by driven-dissipation, exploiting the emission into a one-dimensional photonic bath. In our experiment, a pair of superconducting qubits serve as remote quantum nodes, which are connected via an open microwave waveguide on a chip. In following a previous theoretical proposal \cite{10.1103/physreva.91.042116}, our approach offers simplicity by relying on classical drives and conventional bidirectional qubit-photon couplings. The formation of a dark state in our experiment is achieved via the precise tuning of the qubit transition frequencies, ensuring that the wavelength of the photons emitted by the qubits is precisely matched to the inter-qubit physical distance. Supplying continuous drives through the waveguide, we demonstrate the stabilization of a dark state in this system, which, asymptotically, takes the form of a Bell state under strong drives. 
Following the stabilization, we conduct quantum state tomography to quantify the degree of entanglement in the steady state of the system. Through repeated experiments with different settings, we find a maximum concurrence of 0.504$^{+0.007}_{-0.029}$ (95\% confidence interval) with a stabilization time constant of 56 $\pm$ 4 ns. Moreover, we study the trade-off between the stabilization time and the entanglement quality in the steady state and compare it with a numerical model that considers the noise sources in our experiment. Finally, we discuss future improvements in the experimental parameters, where we expect to reach fidelities exceeding 90$\%$ with improved thermalization of the waveguide and coherence parameters within the reach of state-of-the-art superconducting qubits. By demonstrating driven-dissipative entanglement stabilization in a waveguide, our experiment marks an important step toward realizing a modular network architecture where two- and multi-node remote entanglement is accessed on demand via an open radiation channel.

\subsection*{Theoretical concept}\label{sec:theoretical_concept}

Our system includes two qubits coupled to a shared waveguide with equal dissipation rates of $\Gamma_\mathrm{1D}$ (see \cref{figure 1}a). The qubit frequencies are offset symmetrically with respect to a center frequency ($\omega_\mathrm{1,2} = \omega \pm \delta$). Further, we choose the center frequency $\omega$ such that $\ell = m\lambda$, where $\ell$ is the physical distance between the qubits, $\lambda$ is the wavelength of radiation at $\omega$, and $m$ is an integer. We assume the qubits are driven via a classical coherent field at the frequency $\omega$, supplied through the waveguide. In a frame rotating at the driving frequency, and after applying the rotation wave approximation (RWA), the Hamiltonian for this system can be written as 
\begin{equation}
\hat{H}/\hbar = \sum_{i=\mathrm{1,2}} \frac{\delta_{i}}{2}\hat{\sigma}_{z,i} + \frac{1}{2}\left(\Omega\hat{\sigma}^{\dagger}_{i} + \Omega^{*}\hat{\sigma}_{i}\right).
\label{eq:qubit-basis-hamiltonian}
\end{equation}
Here, $\Omega$ denotes the  Rabi frequency of the drive and $\delta_\mathrm{1,2} = \pm \delta$. While this Hamiltonian is separable, the interference of photons emitted by the two qubits in this setting gives rise to suppression and enhancement of spontaneous emission \cite{Lalumiere:2013io}. This can most easily be understood in the basis of the collective states corresponding to a pair of maximally entangled states, denoted by the triplet and singlet states $|{T,S} \rangle = (|{eg}\rangle \pm  |{ge}\rangle)/\sqrt{2}$. The Hamiltonian in the basis of these states can be written as
\begin{equation}
\hat{H}/\hbar = \frac{\Omega}{\sqrt{2}} \left(| T\rangle \langle gg |+| ee\rangle \langle T |\right) - \delta \left(| S\rangle \langle T |\right)+\mathrm{H. c.}.
\label{eq:ST-basis-hamiltonian}
\end{equation}

\Cref{figure 1}b (left) shows the corresponding energy level diagram, including the coupling and dissipation terms (see \cref{appendix:ME} for derivation). As evident, the singlet state is sub-radiant and protected from direct dissipation into the waveguide. However, in the presence of frequency detuning, the singlet state can exchange population with the waveguide through coherent interaction with the triplet state, which is super-radiant. In the steady state, the combination of this process and continuous drive through the waveguide results in the formation of a superposition of the singlet and the ground states that is dark to the waveguide. This stationary state can be found by solving for an eigenstate of the Hamiltonian that satisfies the condition $(\hat{\sigma}_{1}+\hat{\sigma}_{2}) |D \rangle = 0$ (see \cref{appendix:dark-state-theory} and \cite{10.1103/physrevlett.113.237203, 10.1103/physreva.91.042116}). The diagonalized level structure is shown in \cref{figure 1}b (right), where the dark state is given by
\begin{equation}
|{D}\rangle= \frac{ |{gg}\rangle + \alpha |{S}\rangle}{\sqrt{1+{|\alpha|}^2}}.
\label{eq:dark-state}
\end{equation}
Here, $\alpha =\Omega/\sqrt{2}\delta$ is a drive-power dependent parameter that sets the singlet fraction ($|\alpha|^2/(1+|\alpha|^2)$). As evident, the stationary state is pure, and for strong drives ($|\alpha| \gg 1$) approaches the maximally-entangled singlet state. As a result, by simply supplying the drive into the waveguide, one can stabilize entanglement between the qubits starting from an arbitrary initial state.

By redrawing the energy level diagram to include the dark state, $|{D}\rangle$, and a state normal to it ($|{B}\rangle =  (\alpha|{gg}\rangle -  |{S}\rangle)/{\sqrt{1+{|\alpha|}^2}}$), one can find the rate of pumping population into $|{D}\rangle$. As shown in \cref{figure 1}b (right), the population is dissipatively transferred from $|{T}\rangle$ to $|{D}\rangle$, where it is trapped due to the decoupling of the dark state from the remainder of the energy levels. The effective rate of this process can be found as (see \cref{appendix:dark-state-theory} for derivation) 
\begin{equation}
\gamma_{\mathrm{eff},D} = \frac{2\Gamma_\mathrm{1D}}{1+{\Omega^2/2\delta^2}}.
\label{eq:dark-rate-main}
 \end{equation}
The reciprocal of this `pumping' rate sets the timescale for stabilization $t_D = 1/\gamma_{\mathrm{eff},D} $. We highlight the competition between the singlet fraction $|\alpha|^2/(1+|\alpha|^2)$ and the stabilization time  $t_D = (1+|\alpha|^2)/(2\Gamma_\mathrm{1D})$, where achieving larger singlet fractions - corresponding to more entanglement - requires higher drive powers and longer stabilization times.  (see \cref{appendix:dark-state-theory} for further discussion).

To successfully implement the stabilization protocol under consideration, a physical platform must fulfill several requirements. Most importantly, it is necessary to have precise control over the qubit transition frequency and to establish efficient interfaces between qubit and propagating photons. Precise control over the qubit frequency is needed to ensure that the interference of emitted photons leads to destructive interference at both the outputs of the waveguide simultaneously, culminating in a dark state. Formally, this condition can be articulated by defining the null spaces for the collective jump operators (see \cref{appendix:dark-state-theory}) \footnote{This condition can be relaxed at the expense of realizing chiral emission into the waveguide \cite{10.1088/1367-2630/14/6/063014, 10.1103/physrevlett.113.237203, Rabl2023, 10.1103/physreva.91.042116}}. Efficient qubit-photon interfaces are vital to minimize photon loss during the emission and re-absorption processes among the qubits, which can result in a reduced fidelity for the stabilized state. A key metric in evaluating this effect is the Purcell factor, defined as the ratio of an individual qubit's decay rate to the waveguide to its intrinsic decoherence rate, $P_\mathrm{1D} = \Gamma_\mathrm{1D}/\Gamma'$, where $\Gamma^{'} = 2\Gamma_2 - \Gamma_\mathrm{1D} = \Gamma_\mathrm{int}+2\Gamma_\phi$. ($\Gamma_2$ is the total qubit decoherence, $\Gamma_\mathrm{int}$ is loss to non-radiative channels, and $\Gamma_\phi$ is pure dephasing). In addition to the factors mentioned, another key ingredient is the characterization of the stabilized joint qubit state. This task is particularly challenging in the presence of dissipation from the waveguide. To overcome this challenge, characterization measurements need to happen either on very short time scales, or, alternatively, temporary elimination of waveguide dissipation is needed during the characterization process. In the next section, we detail an experimental realization based on transmon superconducting qubits that satisfies these requirements.

 \subsection*{The experiment}

The fabricated superconducting circuit used to realize our experiment is shown in \cref{figure 1}c. The circuit consists of three transmon qubits (1, 2, and 3), which are side-coupled to the same coplanar waveguide (CPW). Each qubit has a weakly coupled charge control line (shown in orange) and external flux bias port (shown in green) for tuning its transition frequency. Qubit 3 does not participate in any of our experiments and is decoupled from the rest of the system by tuning its frequency well away ($>$1 GHz) from the other two qubits. We list the details of our device fabrication in \cref{appendix:Methods}.

\begin{figure*}[htbp]
\centering
\includegraphics[width=0.85\linewidth]{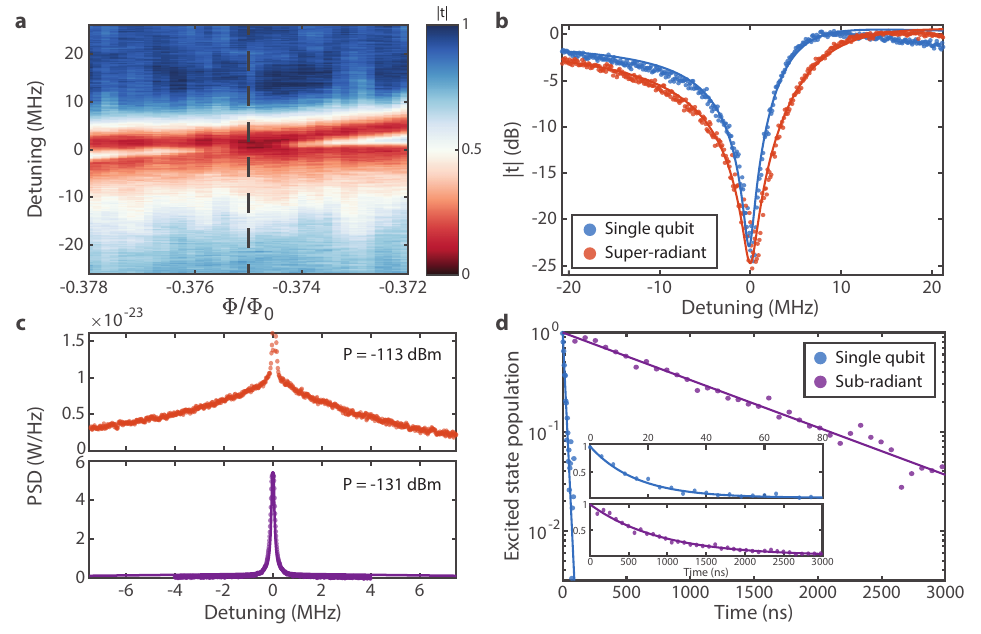}

\caption{\textbf{Characterizing the super- and sub-radiant collective states.} a) Transmission spectrum measured through the waveguide as Qubit 1 is frequency-tuned across Qubit 2. The dashed line denotes the $\ell \approx \lambda$ point). b) Transmission spectra for the single qubit (Qubit 2, blue) and two-qubit (red) settings, where a broader spectrum indicates the super-radiant (triplet) state.  Fitting Lorentzian lineshapes to the two data sets gives decay (intrinsic decoherence) rate rates of $\Gamma_{1\mathrm{D}}/2\pi = 10.7 \pm 0.4$ MHz ($\Gamma^{'}/2\pi = 1.0 \pm 0.5$ MHz) and $18.3 \pm 0.4$  MHz ($\Gamma^{'}/2\pi = 1.3 \pm 0.5$ MHz). The decay (intrinsic decoherence) rate for Qubit 1, tuned to have the same transition frequency, is found as $\Gamma_{1\mathrm{D}}/2\pi = 10.3 \pm 0.3$ MHz ($\Gamma^{'}/2\pi = 0.9 \pm 0.4$ MHz, not shown). c) Inelastic scattering spectrum. At higher drive powers (upper inset), the lineshape includes contributions from both the triplet and singlet states. Reducing drive power results in a lineshape predominantly set by the singlet state. Fitting using master equation simulations gives individual qubit dephasing of 174 $\pm$ 24 kHz and correlated dephasing of 127 $\pm$ 85 kHz. d) Measured relaxation lifetimes for an individual qubit (Qubit 2) and the singlet state yielding $T_1 = 16$ $\pm$ 1.9 ns and $T_1 = 910$ $\pm$ 47 ns, respectively. The insets show the linear scale plots of the same data.}
\label{figure 2}
\end{figure*}

We first tune the qubit transition frequencies to realize the previously described level structure of \cref{figure 1}b. We then verify the required interference conditions by searching for signatures of the sub-radiant and super-radiant states. \Cref{figure 2}a shows the transmission spectra through the waveguide for weak microwave drives, measured as a function of the flux bias of Qubit 1. Meanwhile, Qubit 2 has its transition frequency fixed at $\omega$ such that the inter-qubit separation along the waveguide equates the corresponding wavelength ($\ell = \lambda$). As the two qubits cross, we note a broader resonant lineshape, which corresponds to the formation of a super-radiant state \cite{Lalumiere:2013io}. \Cref{figure 2}b shows the waveguide transmission spectrum for the case where two qubits are precisely on resonance at $\omega$ (red), and for when Qubit 2 is at $\omega$ while Qubit 1 is tuned out of the measurement window (blue). By fitting Lorentzian lineshapes to these spectra, we find the radiative decay rate of the super-radiant (triplet) state,  $\Gamma_{\mathrm{1D},T}/ 2\pi =  18.3$ MHz, which is nearly twice the single-qubit decay rate [$(\Gamma_{\mathrm{1D}, 1},\Gamma_{\mathrm{1D}, 2}) /2\pi = (10.3,10.7)$ MHz], pointing to the correct phase length between the qubits. 

We characterize each individual qubit-photon interface using fits to the single-qubit spectra at $\omega$, extracting Purcell factors for each qubit [$(P_\mathrm{1D,1},P_\mathrm{1D,2}) = (11.4,10.7)$]. Although the singlet is (ideally) protected from emission to the waveguide, in a realistic system with a finite Purcell factor, it has a finite lifetime. Being sub-radiant, though, it is not visible in the waveguide transmission response. Instead, we measure the inelastic scattering from the qubits using a spectrum analyzer (see \cref{appendix:Methods} for measurement details). While the resonance fluorescence spectrum at higher powers contains contributions from both the super- and sub-radiant states, a measurement done at a sufficiently low drive power is dominated by the response from the sub-radiant (singlet) state, as shown in \cref{figure 2}c \cite{vanLoo2013Dec}. A master equation simulation fit to the inelastic scattering profile confirms the presence of the sub-radiant state (see \cref{appendix:dark-state-fluorescence} for a detailed discussion). Furthermore, we measure the singlet's population decay lifetime by resonantly exciting it via the individual charge lines. In this measurement, we read out the population of the ground state using the state-dependent transmission coefficient through the waveguide (similar to \cite{zanner2022}, see \cref{appendix:lifetime}). \cref{figure 2}d shows the measurement results for the singlet state plotted next to the free population decay of an individual qubit (measured using resonant readout, \cref{appendix:lifetime}). We find a large contrast (factor of over 56) in the lifetimes of the single qubit and sub-radiant states, indicating that the qubit frequency configuration and Purcell factors are suitable for the realization of the stabilization protocol.

\begin{figure*}[htbp]
\centering
\includegraphics[width=\linewidth]{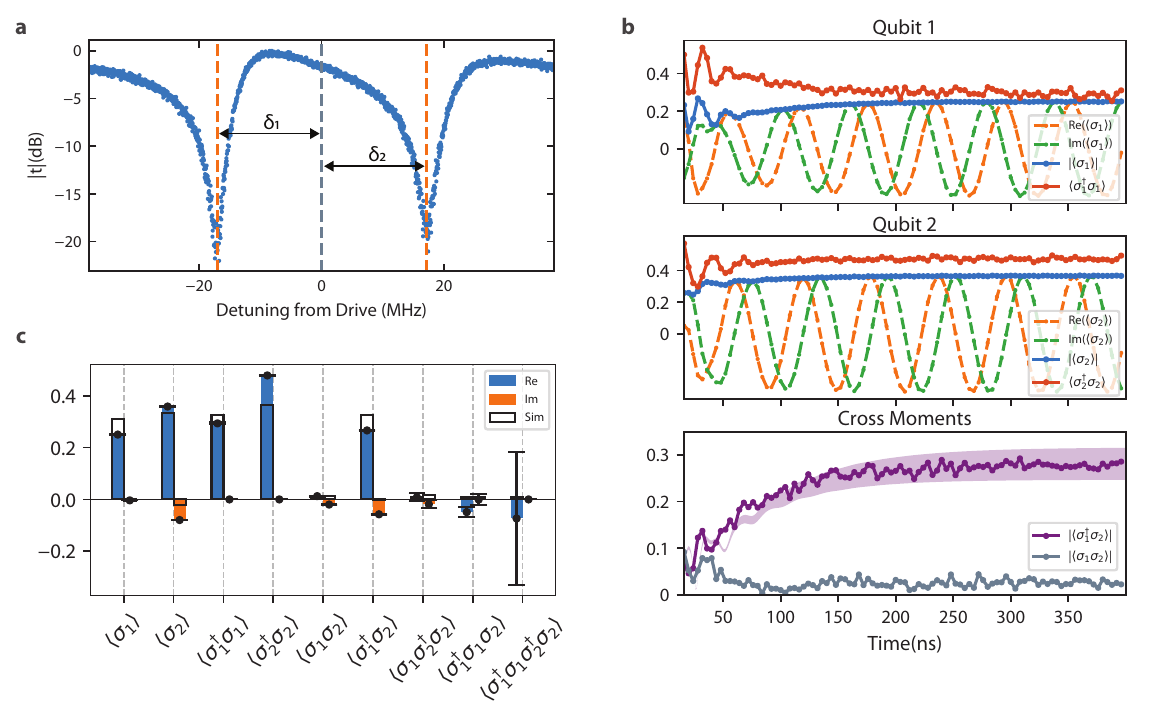}
\caption{\textbf{Stabilization dynamics.} a) The transmission spectrum measured through the waveguide when the two qubits are detuned symmetrically ($\delta/2\pi = 17$ MHz). The black dashed line marks the target drive frequency ($\omega/2\pi = 6.392$ GHz, where $\ell \approx \lambda$). b) The time dynamics of the first- and second-order moments after supplying a continuous drive with a power of $-117$ dBm (corresponding to $\Omega_1/2\pi = 36$ MHz and $\Omega_2/2\pi = 37$ MHz) and frequency of $\omega$ to the waveguide. The moments are measured via the resonant readout of the two-qubit state as described in the main text. The `time' axis on all plots shows the duration for which the drive has been applied to the waveguide. A delay following each measurement (200 ns, much longer than the radiative lifetime of the individual qubits) ensures that the system is at `rest' prior to the subsequent measurement. The plots display average values derived from an experiment that has been repeated 150 million times. c) The measured moments after a long drive (396 ns), corresponding to the system at the steady state. The plot includes all non-zero terms for an arbitrary state residing in a Hilbert space containing at most a single excitation in each qubit. The plots display average values derived from an experiment that has been repeated 3 billion times, and horizontal bars indicate standard deviations (see \cref{appendix:MLE}). The values from a numerical simulation are displayed for comparison (boxed outlines). The model uses parameters based on the single-qubit characterization data. For further simulation details, see  \cref{appendix:simulation}.}
\label{figure 3}
\end{figure*}

Having established the operation frequency from the characterization experiments detailed above, we proceed with the stabilization and characterization of entangled states. In this step, the qubits are detuned with respect to the target frequency (see \cref{figure 3}a). Starting with a system at rest in this setting, we apply a narrow band drive at $\omega$ to initiate the stabilization process. After supplying the drive for a finite duration of time, we turn it off and allow the qubits to freely decay into the waveguide. {We note here that strong engineered dissipation appears to directly preclude characterization of the qubit state; in typical circuit QED settings, qubit state characterization favors preserving the qubit state and minimizing qubit-environment coupling. A natural alternative is therefore to recover the qubit state using the emitted photons.} Our choice of frequency detuning between the two qubits results in a low overlap between their spectral content. Hence, the spontaneous emission from the two qubits into the waveguide has spectrally distinguishable microwave photons with well-defined temporal wavepackets. Using quadrature amplitude detection and mode-matching to the photonic time bins (following previous work, \cite{experimental_state_tomography, 10.1038/nphys2612,10.1038/s41534-020-0266-4,10.1126/sciadv.abb8780}) we successfully measure the self and cross-moments of the emitted photons. Through input-output relations of multiple qubits coupled to an open waveguide, we can relate these photonic moments to those of the qubits themselves. Additionally, since each photon is emitted by a qubit it may at most have only one excitation, we can compute all the self and cross-moments of the form $\langle (\hat{\sigma}_{1}^{\dag})^{n_{1}}\hat{\sigma}_{1}^{m_{1}}(\hat{\sigma}_{2}^{\dag})^{n_{2}}\hat{\sigma}_{2}^{m_{2}} \rangle$  $ \forall$ $ n_{1},n_{2},m_{1},m_{2} \in \{0,1\}$, which suffice to perform a joint state tomography of the two-qubit state (see \cref{appendix:tomography-all}). 

\begin{figure}[htbp]
\centering
\includegraphics[width=0.95\columnwidth]{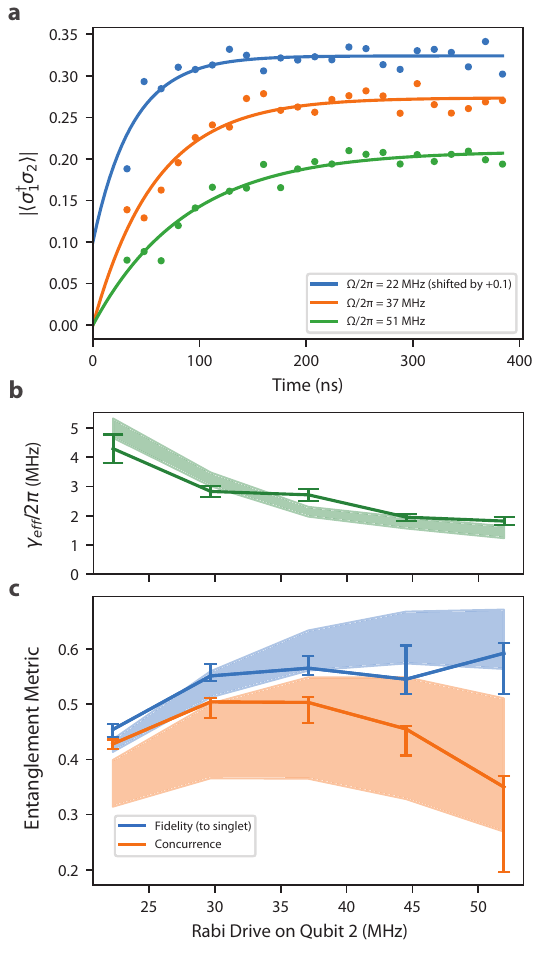}
\caption{\textbf{Interplay between drive power, stabilization rate, and entanglement} a) Stabilization dynamics for different drive powers. The graph shows the $\langle \hat{\sigma}_1^\dagger \hat{\sigma}_2\rangle$ cross moment, which is related to the amount of entanglement in the system. We fit exponential curves (solid lines) to the measured data (dots) and extract the stabilization rates, given by $\gamma_\mathrm{eff}$, for different drive powers. b) Variation in stabilization rates with power. Error bars represent $\pm$ standard deviations. We can see a clear decrease in the stabilization rate with increasing drive power, as expected from the theory (\cref{eq:dark-rate-main}). c) Fidelities (to a singlet state) and concurrence against vs the drive power. Error bars represent 95\% confidence interval. The maximum concurrence achieved is 0.504$^{+0.007}_{-0.029}$ (95\% confidence interval). The corresponding time constant is 56 $\pm$ 4 ns. The decrease in concurrence after a point is explained by a finite Purcell factor. Shaded regions in panels (b) and (c) represent simulations results with varying Purcell factors (from 10 to 30) for the qubits. See \cref{appendix:simulation} for further discussion on simulations.} 
\label{figure 4}
\end{figure}

\Cref{figure 3}b shows the measured self- and cross-moments of the qubits as the state evolves in time for a Rabi drive of $\Omega /2\pi = 37 $ MHz and a detuning of $\delta/2\pi=17$ MHz between the qubits and the drive. As evident, the system starts with the qubit pair in the ground state, where all the moments are zero. With the continuous drive turned on, we observe an initial increase in the average population of each qubit with time, followed by settling into a steady-state value. The stabilization of a dark state with entanglement can be identified via the emergence and stabilization of a non-zero cross-qubit moment $\langle \hat{\sigma}_1^\dagger \hat{\sigma}_2\rangle$. We also note the nonzero expectation values of the single-qubit first moments $\langle\hat{\sigma}_i\rangle$, which can be attributed to the finite population of the ground state. \Cref{figure 3}c plots the summary of the measurement results of the steady state and includes all relevant moments for any joint state of two qubits. The small magnitudes of the third and fourth moments indicate the low population of the doubly excited state $|ee\rangle$. Using the computed moments we reconstruct the density matrix for the two qubit system using maximum-likelihood estimation (\cref{appendix:MLE}). The fidelity of the extracted density matrix to the singlet state at this drive power and detuning is 56.5\% (95\% confidence interval from 55.2\% to 58.7\%), which is already above the 50\% threshold that confirms entanglement.

Beyond the quality of entanglement, the stabilization rate is another important metric for practical applications.  In the protocol being examined, the stabilization rate is anticipated to decrease as the drive power increases, as indicated by \cref{eq:dark-rate-main}. At the same time, the level of entanglement, quantified by the singlet fraction (see \cref{eq:dark-state}), is expected to rise with increasing drive power. \Cref{figure 4}a shows the time evolution of the cross moment $\langle \hat{\sigma}_1^\dagger \hat{\sigma}_2\rangle$  across three different power levels. The stabilization rates, derived from exponential fits to this data (and experiments at a few different drive powers), exhibit a consistent decline with increased power, aligning with our expectations (see \cref{figure 4}b). We also plot the fidelity relative to the singlet state for each power setting (see \cref{figure 4}c), revealing that fidelities initially rise with power before reaching a plateau.

To more accurately assess entanglement, we calculate and graph the concurrence \cite{PhysRevLett.80.2245} for each power setting (see \cref{figure 4}c). This graph presents a more distinct trend, showing concurrence peaking at intermediate drive powers and diminishing at both lower and higher powers. The reduction of entanglement at low powers aligns with expectations, mirroring the calculated fidelity and our model's predictions. The reduction in entanglement at high power can be qualitatively understood by considering the role of parasitic (i.e., non-radiative) damping and dephasing, which our analytical model does not capture. Intuitively, these effects drive the qubits towards a mixed state, with the decoherence rate  $\Gamma'$ competing against the effective pumping rate $\gamma_{\mathrm{eff}}$ (see \cite{10.1088/1367-2630/14/6/063014}). This explanation is supported by calculating the purity of the stabilized states, which drop from 90\% at $\Omega/2\pi = 30$ MHz to 60\% at $\Omega/2\pi = 51$ MHz. Since our analytical model does not account for parasitic damping and dephasing, we utilize master equation simulations to model our experimental setup more accurately.  A complicating factor for these simulations is power-dependent dephasing in driven qubits, as documented in previous research \cite{Yan2013Aug, Ithier2005Oct}, which effectively results in a variable Purcell factor at different drive powers  (see \cref{appendix:noise}). To capture this phenomenon, we conduct master equation simulations of our experiment for a range of Purcell factors, setting the lower limit based on our qubit spectrum measurements ($P_\mathrm{1D}= 10$) and selecting an upper bound that qualitatively aligns with our experimental data ($P_\mathrm{1D}= 30$). The simulations, depicted in \cref{figure 4}b and c, exhibit trends that closely mirror the observed power-dependent variations in our experimental results. In our system, we can achieve a maximum concurrence of 0.504$^{+0.007}_{-0.029}$ (95\% confidence interval) at an optimal drive power of $\Omega/2\pi = 30$ MHz. 

\subsection*{Conclusions and outlook}

Remote entanglement is a crucial resource for quantum networks with a wide range of applications in distributed quantum computing, communication, and sensing. In platforms equipped with low-loss communication channels, like superconducting qubits, remote entanglement has been produced through deterministic processes involving direct photon exchange \cite{10.1038/s41586-023-05885-0, Kurpiers_2018, PhysRevLett.120.200501, Axline_2018} or virtual interactions via cavity relays \cite{10.1038/s41567-019-0507-7, leung_deterministic_2019, Zhong2021Feb, Niu2023Mar}. Although these methods have achieved high fidelities, they typically require pulse shaping or precise timing of qubit operations. Relying on driven-dissipative stabilization, as showcased in our experiment, offers a simple alternative that relaxes these requirements. Additionally, stabilizing the target state allows for maintaining the entanglement indefinitely, until it is needed for quantum networking operations, thereby providing an on-demand resource without the latency caused by the propagation delay between the nodes.

In terms of physical dimensions, the inter-qubit distance in our experiment is not particularly long (all components are confined within a 1 cm by 1 cm chip) and is comparable to previous work in cavity systems
\cite{10.1038/nature12802,10.1103/physrevlett.116.240503,10.1103/physrevx.6.011022,10.1038/s41467-022-31638-0}. Distinctly, however, relying on an open waveguide makes the concept insensitive to the distance between the qubits (within integer multiples of $\lambda/2$), allowing for the stabilization of long-distance entanglement.  This increased range of entanglement is not limitless, as the breakdown of the Markov approximation (inherently assumed in our analysis) sets an upper bound
on the distance. Nonetheless, with the dissipation and drive rates attainable with superconducting
qubits, this limit is rather long (meters), allowing
for the method to be practically feasible for a range of
applications. Going beyond this limit remains relatively unexplored, though proposals based on injecting squeezed light into waveguides may provide a route towards this \cite{Rabl2023,10.1103/physrevresearch.4.023010}. Additionally, the waveguide's broad-band nature enables frequency multiplexing, permitting the simultaneous entanglement of multiple qubit pairs at different frequencies through the same channel. 
Finally, using chiral qubit-photon coupling \cite{sollner2015, scheucher2016, Kannan2022Mar, 10.1038/s41467-023-38761-6,10.1103/physrevx.13.021039} can relax the requirement of precise phase tuning between the qubits and also lead to the formation of multi-partite entanglement \cite{10.1103/physreva.91.042116, 10.1103/physrevlett.113.237203}.  

In terms of entanglement fidelity, we achieve only modest results, which are limited by the qubit coherence (see \cref{appendix:achievable_fidelities}). However, with improvements to qubit coherence and reduced waveguide temperature, we expect fidelities of over 90\% (concurrence of 0.88) to be within the reach of this method. In terms of bandwidth, the stabilization rate of 2.83 ($\pm$0.19) MHz (settling time of 56 ns) is comparable to previous demonstrations with superconducting qubits \cite{PhysRevX.6.031036, Kurpiers_2018, leung_deterministic_2019, 10.1038/s41567-019-0507-7}. We note that predicted fidelity improvements are expected to be accompanied by a reduction in stabilization times (a trend that goes beyond our experiment and is typical of driven-dissipative processes). Nevertheless, for fidelities over 90\% we expect a stabilization rate of about 250 kHz to be within reach. Ultimately, going beyond these limits requires further investigations of protocols with better hardware efficiency and improved bandwidths \cite{Lingenfelter2023Jul}.

In conclusion, we have demonstrated the generation and stabilization of entanglement between a pair of distant qubits coupled to a shared waveguide via a driven-dissipative protocol. The presented scheme offers simplicity, steady state operation, and decoherence protection, making it an attractive avenue for remote entanglement generation. We have also identified the limitations of our experiment, finding high-fidelity remote entanglement to be achievable using qubits with improved decoherence characteristics well within reach of existing superconducting technologies. With future improvements in fidelities, we envision that entanglement stabilization protocols may find applications in distributed quantum computing and quantum communication. Beyond applications, extending the range of entanglement stabilization may prompt studies of non-local noise in photonic reservoirs, which may lead to potentially relevant insights for quantum error correction.

\section*{acknowledgments}
This work was supported by startup funds from Caltech's EAS division, National Science Foundation (award number:\,1733907), and Office of Naval Research (award number:\, N00014-24-1-2052).  P.S.S. gratefully acknowledges support from the S2I-Gupta Fellowship. F.Y. gratefully acknowledges support from the NSF Graduate Research Fellowship. C.J. gratefully acknowledges support from the IQIM/AWS Postdoctoral Fellowship.
 
\bibliography{references}
\clearpage
\appendix
\section{Methods}
\label{appendix:Methods}
\subsection{Fabrication}
Our device is fabricated on a 1 cm $\times$ 1 cm high-resistivity (10 k$\Omega$-cm) silicon substrate. Electron-beam lithography is used to pattern the structures in separate metal layers on the chip. Each lithography step is followed by electron-beam evaporation of metal and liftoff in N-methyl-2-pyrrolidone at 150$^\circ$ C for 1.5 hours. Device layers are as follows. (i) 150 nm thick niobium markers, deposited at 3 \text{\normalfont\AA}/s. (ii) 120 nm thick aluminum ground plane, waveguide, flux lines, and qubit capacitors, deposited at 5 \text{\normalfont\AA}/s. (iii) Josephson junctions evaporated (at 5 \text{\normalfont\AA}/s) using double angle evaporation and consisting of 60 nm and 120 nm layers of aluminum, with 15 minutes of static oxidation between layers. We use asymmetric Josephson junctions in the SQUID loop of each qubit to mitigate the effects of dephasing \cite{Hutchings2017Oct}. For further details, see device parameters given in \cref{tab:experimental_params}. (iv) 150 nm thick aluminum band-aids and air-bridges, deposited at 5 \text{\normalfont\AA}/s. Band-aids ensure electrical contact between Josephson junctions and qubit capacitors. Air bridges are used to ensure the suppression of the slot-line modes in the waveguide \cite{Chen2014Feb}. Air-bridges are patterned using grey-scale electron-beam lithography and developed in a mixture of isopropyl alcohol and de-ionized water, followed by 10 minutes of reflow at 105$^\circ$ C \cite{Painter2020Dec}. Electron beam evaporation of the band-aid/bridge layer is preceded by 7 minutes of Ar ion milling.

\begin{figure}[htbp]
\centering
\includegraphics[width=1\linewidth]{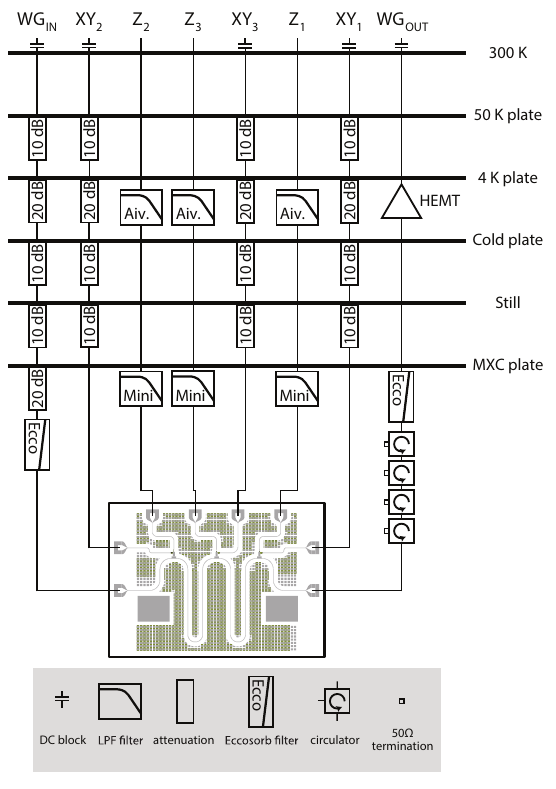}
\caption{\textbf{Measurement Setup} schematic of dilution fridge wiring for measurement.}
\label{fig:Fridge}
\end{figure}

\subsection{Measurement setup}
Measurements are performed in a $^3$He/$^4$He dilution refrigerator. A schematic of our measurement setup is shown in \cref{fig:Fridge}. The fabricated chip is wire-bonded to a PCB and placed in an copper box. The box is then mounted to the mixing plate which is cooled to a base temperature of 10 mK. 

The waveguide input line ($\mathrm{WG_{\mathrm{IN}}}$) is used to drive qubits and is attenuated at each temperature stage to minimize thermal noise; the total attenuation is 70 dB. Attenuators (not shown) are added to the input line at room temperature to control input power. Four isolators are used to reduce thermal noise in the waveguide output line ($\mathrm{WG_{\mathrm{OUT}}}$). The output is amplified by a high electron mobility transistor (HEMT) amplifier at the 4 K stage and a room temperature amplifier (not shown) outside of the fridge.

A low noise, multi-channel DC source provides current biases to flux tune qubit frequencies. Low-pass RF filters (Aivon Therma-uD25-GL RC filter with 15 kHz cutoff frequency, Mini-Circuits VLFG490+ with 490 MHz cutoff frequency) suppress high-frequency thermal noise in the DC lines, which are not attenuated. Qubits are also equipped with XY drive lines; RF inputs to the drive lines are attenuated (50 dB total) to reduce thermal noise. 

\paragraph*{Spectral measurements}
Elastic scattering measurements (\cref{figure 2}a,b; \cref{figure 3}a) are performed using a vector network analyzer (VNA, Agilent N5242A). VNA drives are attenuated to sub-single-photon power levels in order to prevent qubit saturation. The non-linear least squares method or circle fit method is used to fit transmission traces. Inelastic scattering measurements (\cref{figure 2}c) are performed with an RF spectrum analyzer (Rohde and Schwarz FSV3013); the microwave excitation tone is provided by the VNA in zero-span mode. The spectrum analyzer acquisition is performed with a resolution bandwidth of 20 kHz. Measured resonance fluorescence of the sub-radiant state is fitted to master equation simulations, as discussed in \cref{appendix:dark-state-fluorescence}.

\paragraph*{State tomography} 
Qubit state tomography is performed using the Quantum Machines OPX+ (QM) module, which is capable of arbitrary waveform generation and heterodyne detection. To generate the drive, MHz frequency intermediate frequency (IF) signals from the QM are combined with a local oscillator (LO) supplied by an RF signal generator (Rohde and Schwarz SMB100) using IQ mixers (Marki Microwave MMIQ-0520LS). For readout, the signal from the output line is downconverted using an IQ mixer, and the resulting IF signal is digitally demodulated. Details of state tomography are discussed in \cref{appendix:tomography}. Data presented in \cref{figure 2}d; \cref{figure 3}b, c; and \cref{figure 4}a, b are recorded using state tomography techniques.

\subsection{Device parameters}
\label{appendix:table}
Relevant device parameters are given in \cref{tab:experimental_params}.

\begin{table}[h]
\centering
\resizebox{1\linewidth}{!}{
\begin{tabular}{lcr}
\toprule

Description & Symbol & Value \\
\midrule
Qubit 1  \\
\midrule
Minimum frequency & $\omega_\mathrm{1, min}/2\pi$ & 6.409 GHz \\
Maximum frequency & $\omega_\mathrm{1, max}/2\pi$ & 7.505 GHz \\
Total Josephson energy & $E_\mathrm{J}$ $(= E_\mathrm{J,a}+E_\mathrm{J,b})$ & 25.4 GHz \\
Junction asymmetry & $\alpha_\mathrm{J}$ $(= E_\mathrm{J,a}/E_\mathrm{J,b})$ & 6.6 \\
Anharmonicity & $E_\mathrm{c}$ ($= -\alpha$) & 300 MHz \\

\midrule
Qubit 2 \\
\midrule
Minimum frequency & $\omega_\mathrm{2, min}/2\pi$ & 6.343 GHz \\
Maximum frequency & $\omega_\mathrm{2, max}/2\pi$ & 7.457 GHz \\
Total Josephson energy & $E_\mathrm{J}$ $(= E_\mathrm{J,a}+E_\mathrm{J,b})$ & 25.1 GHz \\
Junction asymmetry & $\alpha_\mathrm{J}$ $(= E_\mathrm{J,a}/E_\mathrm{J,b})$ & 6.5 \\
Anharmonicity & $E_\mathrm{c}$ ($= -\alpha$) & 300 MHz \\

\midrule
Qubit 3 \\
\midrule
Minimum frequency & $\omega_\mathrm{3, min}/2\pi$ & 6.565 GHz \\
Maximum frequency & $\omega_\mathrm{3, max}/2\pi$ & 7.640 GHz \\
Total Josephson energy & $E_\mathrm{J}$ $(= E_\mathrm{J,a}+E_\mathrm{J,b})$ & 26.3 GHz \\
Junction asymmetry & $\alpha_\mathrm{J}$ $(= E_\mathrm{J,a}/E_\mathrm{J,b})$ & 6.9 \\
Anharmonicity & $E_\mathrm{c}$ ($= -\alpha$) & 300 MHz \\

\midrule
Super (sub)-radiant state (\cref{figure 2}) \\
\midrule
Waveguide distance & $\ell$ $(=\lambda)$ & 18.2 mm \\
Qubit 1 waveguide decay & $\Gamma_\mathrm{1D,1}/2\pi$ & 10.3 $\pm$ 0.3 MHz \\
Qubit 1 intrinsic decoherence & $\Gamma^\prime_{1}/2\pi$ & 0.9 $\pm$ 0.4 MHz \\
Qubit 1 Purcell factor & $P_\mathrm{1D,1}$ $(=\Gamma_\mathrm{1D,1}/\Gamma^\prime_{1})$ & 11.4 \\
Qubit 2 waveguide decay & $\Gamma_\mathrm{1D,2}/2\pi$ & 10.7 $\pm$ 0.4 MHz \\
Qubit 2 intrinsic decoherence & $\Gamma^\prime_{2}/2\pi$ & 1.0 $\pm$ 0.5 MHz \\
Qubit 2 Purcell factor & $P_\mathrm{1D,2}$ $(=\Gamma_\mathrm{1D,2}/\Gamma^\prime_{2})$ & 10.7 \\

Super-radiant state waveguide decay & $\Gamma_{\mathrm{1D},T}/2\pi$ & 18.3 $\pm$ 0.4 MHz \\
Super-radiant state intrinsic decoherence & $\Gamma^\prime_{T}/2\pi$ & 1.3 $\pm$ 0.5 MHz \\
Qubit 1 lifetime & $\mathrm{T}_{1,1}$ & 16.6 $\pm$ 0.6 ns \\
Qubit 2 lifetime & $\mathrm{T}_{1,2}$ & 16.0 $\pm$ 1.9 ns \\
Sub-radiant state lifetime & $\mathrm{T}_{1,S}$ & 910 $\pm$ 47 ns \\ 
\midrule
Stabilization experiment (\cref{figure 3}) \\
\midrule
Drive frequency & $\omega/2\pi$ & 6.392 GHz \\
Qubit 1 waveguide decay & $\Gamma_\mathrm{1D,1}/2\pi$ & 8.7 $\pm$ 0.1 MHz \\
Qubit 1 intrinsic decoherence & $\Gamma^\prime_{1}/2\pi$ & 1.5 $\pm$ 0.2 MHz \\
Qubit 1 Purcell factor & $P_\mathrm{1D,1}$ $(=\Gamma_\mathrm{1D,1}/\Gamma^\prime_{1})$ & 5.8 \\

Qubit 2 waveguide decay & $\Gamma_\mathrm{1D,2}/2\pi$ & 10.5 $\pm$ 0.1 MHz \\
Qubit 2 intrinsic decoherence & $\Gamma^\prime_{2}/2\pi$ & 1.3 $\pm$ 0.2 MHz \\
Qubit 2 Purcell factor & $P_\mathrm{1D,2}$ $(= \Gamma_\mathrm{1D,2}/\Gamma^\prime_{2})$ & 8.1 \\
\bottomrule
\end{tabular}
}  
\caption{Summary of experimental parameters. }
\label{tab:experimental_params}
\end{table}

\section{Theoretical modeling}
\label{appendix:theory}

Here, we present the theoretical framework used to model two driven qubits coupled to a waveguide and the stabilization of the entangled dark state. 

\subsection{Master equation}
\label{appendix:ME}
The most general master equation for this system is given below \cite{Lalumiere:2013io, 10.1103/physreva.91.042116, Gheeraert2020Nov}.

\begin{align}
\dot{\rho} = -\dfrac{i}{\hbar}[\hat{H} + \hat{H}_{\mathrm{int}} , \rho] + \mathcal{L}\rho
\label{eq:master-equation}
\end{align}

\noindent where 

\begin{equation}
\hat{H}/\hbar = \sum_{i=\mathrm{1,2}} \frac{\delta_{i}}{2}\hat{\sigma}_{z,i} + \frac{1}{2}\left(\Omega_{i}\hat{\sigma}^{\dagger}_{i} + \Omega^{*}_{i}\hat{\sigma}_{i}\right).
\label{eq:qubit-basis-hamiltonian-sup}
\end{equation}

\noindent is the system Hamiltonian written in the drive frame after applying the rotating wave approximation (RWA). Both qubits are driven at the same frequency $\omega$ such that $\delta_{i} = \omega_{i} - \omega$ is the qubit-drive detuning ($\omega_i$ is the qubit $i$ frequency). $\Omega_i$ denotes the Rabi frequency on each qubit. In the case of a drive applied via the waveguide (corresponding to our experiment), $\Omega_i = |\Omega_i|e^{ikx_i}$ includes the propagation phase accumulated by the drive. Here, $x_i$ is the position of qubit $i$ and $k = \frac{n\omega}{c}$ is the wavevector of the drive ($n$ denotes refractive index, $c$ denotes speed of light). $\hat{H}_{\mathrm{int}}$ denotes the photon-mediated interaction.

\begin{equation}
\hat{H}_{\mathrm{int}}/\hbar = J\hat{\sigma}^{\dagger}_{1}\hat{\sigma}_{2} + J^{*}\hat{\sigma}^\dagger_{2}\hat{\sigma}_{1}.
\label{eq:photon-mediated-detuned}
\end{equation}

\vspace*{1px}

\noindent Here, $J = {\sqrt{\Gamma_{\mathrm{1D,2}}\Gamma_{\mathrm{1D,2}}}}{(e^{+ik_{2}d}-e^{-ik_{1}d})}/{4i}$. For the case of two qubits at the same frequency, this expression simplifies to the more familiar $J = ({\sqrt{\Gamma_{\mathrm{1D,2}}\Gamma_{\mathrm{1D,2}}}}/2)\sin(kd)$ \cite{Lalumiere:2013io}. Here, $\Gamma_{\mathrm{1D,i}}$ is the dissipation rate of qubit $i$ into the waveguide, $k_i$ is the wavevector at $\omega_i$, and $d = |x_2 - x_1|$ is the qubit separation. The Liouvillian is given by

\begin{align}
\begin{split}
\mathcal{L}\rho &= \Gamma_{\mathrm{1D,1}}\mathcal{D}[\hat{\sigma}_{1}] \rho + \Gamma_{\mathrm{1D,2}}\mathcal{D}[\hat{\sigma}_{2}] \rho 
\\ & \quad + \Gamma_{12}\mathcal{D} [\hat{\sigma}_{1}, \hat{\sigma}_{2}] \rho + \Gamma_{21}\mathcal{D}[\hat{\sigma}_{2}, \hat{\sigma}_{1}] \rho.
\end{split}    
\label{eq:Liouvillian-full}
\end{align}

Here, $\Gamma_{12} = \Gamma^{*}_{21} = \sqrt{\Gamma_\mathrm{1D,1}\Gamma_\mathrm{1D,2}}({e^{+ik_2d}+e^{-ik_1d}})/{2}$, which simplifies to $\Gamma_{12} = \sqrt{\Gamma_\mathrm{1D,1}\Gamma_\mathrm{1D,2}}\cos(kd)$ for resonant qubits. This term denotes the correlated decay between qubits. The dissipator terms in the equation above are given by

\begin{align}
    \mathcal{D}[A]\rho = A\rho A^\dagger - \dfrac{1}{2} A^\dagger A \rho - \dfrac{1}{2}\rho A^\dagger A .
\end{align}

\begin{align}
    \mathcal{D}[A, B]\rho = B\rho A^\dagger - \dfrac{1}{2} A^\dagger B \rho - \dfrac{1}{2}\rho A^\dagger B .
\end{align}

In our experiment, we drive both qubits through the waveguide at a frequency $\omega$ corresponding to qubit separation of $\lambda$. We then have $\Omega_1 = \Omega_2$.
Additionally, each qubit is equally detuned from the drive frequency so that $\omega_{1,2} = \omega \pm \delta$. At this detuning setting, the photon-mediated interaction ($J$) disappears exactly, and the Hamiltonian is fully described by \cref{eq:qubit-basis-hamiltonian}, reproduced below. 
\begin{equation}
\hat{H}/\hbar = \sum_{i=\mathrm{1,2}} \frac{\delta_{i}}{2}\hat{\sigma}_{z,i} + \frac{1}{2}\left(\Omega\hat{\sigma}^{\dagger}_{i} + \Omega^{*}\hat{\sigma}_{i}\right).
\label{eq:qubit-basis-hamiltonian-sup-simplified}
\end{equation}
Using the triplet and singlet states $|{T,S} \rangle = (|{eg}\rangle \pm  |{ge}\rangle)/\sqrt{2}$, we can then rewrite the Hamiltonian as in \cref{eq:ST-basis-hamiltonian}, repeated below.
\begin{equation}
\hat{H}/\hbar = \frac{\Omega}{\sqrt{2}} \left(| T\rangle \langle gg |+| ee\rangle \langle T |\right) - \delta \left(| S\rangle \langle T |\right)+\mathrm{H. c.}
\label{eq:ST-basis-hamiltonian-sup}
\end{equation}

For this detuning setting, the correlated decay becomes $\Gamma_{12} = \sqrt{\Gamma_\mathrm{1D,1}\Gamma_\mathrm{1D,2}}\exp{(-i\Delta kd)}$, where $\Delta k =  n\delta/c$. The first order correction to correlated decay ($\Delta kd$) is approximately $0.02$ in our experiment and may be safely ignored so that $\Gamma_{12} \approx \sqrt{\Gamma_\mathrm{1D,1}\Gamma_\mathrm{1D,2}}$. (For our experiment, $n = 2.6,$ $\delta = 2\pi\times 17$ MHz $, c = 3\times 10^8$ m/s, $d = 18.2 $ mm.) If we additionally take the assumption that $\Gamma_\mathrm{1D,1} = \Gamma_\mathrm{1D,2}$, the Liouvillian in \cref{eq:Liouvillian-full} simplifies to

\begin{align}
\begin{split}
\mathcal{L}\rho &= \Gamma_{\mathrm{1D}}\mathcal{D}[\hat{\sigma}_{1} + \hat{\sigma}_{2}] \rho
\end{split}    
\label{eq:Liouvillian-qubits}
\end{align}
which may be rewritten straightforwardly as 
\begin{align}
\begin{split}
\mathcal{L}\rho &= 2\Gamma_{\mathrm{1D}}\mathcal{D}[|gg\rangle\langle T| + |T\rangle \langle ee|] \rho
\end{split}    
\label{eq:Liouvillian-simplified-sup}
\end{align}

The simplified Hamiltonian (\cref{eq:ST-basis-hamiltonian-sup}) and Liouvillian (\cref{eq:Liouvillian-simplified-sup}) are illustrated in the left sub-panel of \cref{figure 1}b. The triplet state $|T\rangle$ is super-radiant and decays at rate $2\Gamma_{\mathrm{1D}}$ to the ground state $|gg\rangle$. The singlet $|S\rangle$ has no decay but exchanges population with the triplet at rate $-\delta$. We note that we have considered here the case where qubit separation $d = m\lambda$. For qubit separations of $d = \lambda/2 \pm m\lambda$, the singlet and triplet are exchanged ($|S\rangle$ decays directly to $|gg\rangle$).

\subsection{Dark state formation}
\label{appendix:dark-state-theory}
In our experiment, the interplay between drive, detuning, and decay results in the stabilization of a pure entangled state that is dark to the waveguide. In this section, we detail sufficient conditions for the formation of the dark state and discuss their physical meaning, closely following the references in \cite{10.1088/1367-2630/14/6/063014, 10.1103/physrevlett.113.237203, 10.1103/physreva.91.042116}.

The conditions for the existence of a pure, dark, stationary state $|\Psi\rangle$ are as follows.
\begin{enumerate}
  \item $|\Psi\rangle$ is annihilated by the collective jump operators (or equivalently, $|\Psi\rangle$ exists in the nullspace of the collective jump operators). $c_{\mathrm{R,L}}|\Psi\rangle = 0$. This condition ensures that $|\Psi\rangle$ does not decay to the waveguide output ports and is therefore ``dark."
  \item $|\Psi\rangle$ is an eigenstate of the Hamiltonian. $\hat{H}|\Psi\rangle = \mathrm{E}|\Psi\rangle$. This condition ensures stationarity. 
\end{enumerate}

The collective jump operators for qubits coupled to a waveguide are given by $c_{\mathrm{R}} = \sum_{i}e^{ikx_i}\hat{\sigma}_i$ and $c_{\mathrm{L}} = \sum_{i}e^{-ikx_i}\hat{\sigma}_i$, corresponding to collective decay to the right and left output ports. Here, $k$ is the wavevector at the drive frequency ($k = n\omega/c$) and $x_i$ is the position of qubit $i$. We note that these jump operators are valid in the regime that $\omega \approx \omega_1 \approx \omega_2$. If waveguide dissipation rates $\Gamma_\mathrm{1D}$ are small compared to qubit frequencies ($\omega_1, \omega_2$), and the inter-qubit separation $d$ does not greatly exceed a wavelength, variations in phase shift due to different frequencies is negligible \cite{Kockum2018Apr}. Hence, the interference effects discussed in the following section remain valid.

We first consider condition (1) that $c_{\mathrm{L}}|\Psi\rangle = c_{\mathrm{R}}|\Psi\rangle = 0$. For the case of two qubits coupled to a waveguide, the jump operators may be written as $c_{\mathrm{R}} = \hat{\sigma}_1 + e^{ikd}\hat{\sigma}_2$ and $c_{\mathrm{L}} = \hat{\sigma}_1 + e^{-ikd}\hat{\sigma}_2$, where $d = |x_2 - x_1|$ is the qubit separation. The dissipators corresponding to these operators are $\frac{\Gamma_{\mathrm{1D}}}{2}\mathcal{D}[c_{\mathrm{R,L}}] \rho$. We note that for qubit separation of $d = \lambda$, these dissipators sum to the Liouvillian of \cref{eq:Liouvillian-qubits}. In general, the nullspace of $c_{\mathrm{R,L}}$ consists of $|gg\rangle$ and $|\Psi_{\mathrm{R,L}}\rangle = (|ge\rangle - e^{\pm ikd}|eg\rangle)/\sqrt{2}$. The state $|\Psi_{\mathrm{R,L}}\rangle$ may be interpreted as the collective qubit state that results in destructive interference of waveguide emission in the right (left) direction.
For arbitrary qubit separation $d$, no state other $|gg\rangle$ than satisfies the destructive interference condition in both waveguide directions. However, when $d = m\lambda$ (or $d = \lambda/2 + m\lambda$) for integer m, the singlet (triplet) $|S,T\rangle = (|{ge}\rangle \mp  |{eg}\rangle)/\sqrt{2}$ will be annihilated by both jump operators simultaneously, satisfying (1). We emphasize here the importance of precise control of phase delay between qubits to achieve destructive interference to both waveguide ports. For our experiment, flux-tuning the qubit frequencies allows us to tailor the phase delay in situ. 

While $|gg\rangle$ and $|S\rangle$ annihilate both jump operators for $d = m\lambda$, neither are eigenstates of the Hamiltonian (\cref{eq:ST-basis-hamiltonian}, \cref{eq:ST-basis-hamiltonian-sup}). We note here that for arbitrary qubit detunings ($\delta_i$) and Rabi drive frequencies ($\Omega_i$), it is not possible to find an eigenstate of the Hamiltonian within the considered subspace. However, letting $\Omega_1 = \Omega_2$ and $\delta_1 = -\delta_2$ is sufficient to obtain a ``dark" eigenstate by combining $|gg\rangle$ and $|S\rangle$. This state is given below (and in \cref{eq:dark-state}).

\begin{equation}
|{D}\rangle= \frac{|{gg}\rangle + \alpha |{S}\rangle}{\sqrt{1+{\alpha}^2}} .
\label{eq:dark-state-sup}
\end{equation}

Here, the coherent drive, waveguide dissipation, and detunings all destructively interfere to remove coherent interactions between $|D\rangle$ and the rest of the state space. The parameter $\alpha =\Omega/\sqrt{2}\delta$ sets the the singlet fraction (defined here as $\alpha^2/(1+\alpha^2)$). For large Rabi drives, $|D\rangle$ asymptotically approaches the Bell state $|S\rangle$. By expressing the orthogonal state to $|D\rangle$ in the dark subspace, we may obtain intuition for the population dynamics. This orthogonal state is denoted as $|B\rangle$, and is given below.

\begin{equation}
|{B}\rangle= \frac{\alpha|{gg}\rangle - |{S}\rangle}{\sqrt{1+{\alpha}^2}} .
\label{eq:bright-state-sup}
\end{equation}

We observe that the ground state $|gg\rangle$ may be re-expressed as $|gg\rangle = (|D\rangle + \alpha|B\rangle)/\sqrt{1+\alpha^2}$, allowing us to re-express the Liouvillian in \cref{eq:Liouvillian-simplified-sup} as 

\begin{align}
\begin{split}
\mathcal{L}\rho &= 2\Gamma_{\mathrm{1D}}\mathcal{D}
\left [ \frac{1}{\sqrt{1+\alpha^2}}|D\rangle\langle T|
\right.\\ & \left. 
+ \frac{\alpha}{\sqrt{1+\alpha^2}}|B\rangle\langle T| + |T\rangle \langle ee|  \right ] \rho
\label{eq:Liouvillian-simplified-dark}
\end{split}
\end{align}

\noindent This yields the effective decay (pumping) rates into $|D\rangle$ and $|B\rangle$ from $|T\rangle$.

\begin{equation}
\gamma_{\mathrm{eff},D} = \frac{1}{1+\alpha^2}\times 2\Gamma_\mathrm{1D} = \frac{2\Gamma_\mathrm{1D}}{1+{\Omega^2/2\delta^2}}
\label{dark-rate}
\end{equation}
 
\begin{equation}
\gamma_{\mathrm{eff},B} = \frac{\alpha}{1+\alpha^2}\times 2\Gamma_\mathrm{1D} = \frac{2\Gamma_\mathrm{1D}\Omega}{(1+{\Omega^2/2\delta^2})\delta\sqrt{2}}.
\label{bright-rate}
\end{equation}

\cref{dark-rate} is given in \cref{eq:dark-rate-main} of the main text. Similarly re-expressing the Hamiltonian in \cref{eq:ST-basis-hamiltonian-sup} yields an effective Rabi drive between $|B\rangle$ and $|gg\rangle$ given below.

\begin{equation}
\Omega_{\mathrm{eff},B} = \frac{1}{\sqrt{1+\frac{\Omega^2}{2\delta^2} } } \left ( \frac{\Omega^2+2\delta^2}{2\delta}\right )
\label{bright-rabi}
\end{equation}

These decay and Rabi rates are illustrated in the right sub-panel of \cref{figure 1}b. Here, we note the trade-off between fidelity of the stationary state to $|S\rangle$ (increases with $\alpha/(1+\alpha^2)$) and the decay rate into the dark state (increases with $1/(1+\alpha^2)$). This trade-off is typical of driven-dissipative stabilization schemes \cite{10.1038/s41467-022-31638-0}.

\section{Inelastic scattering of sub-radiant state}
\label{appendix:dark-state-fluorescence}
Resonance fluorescence measurements of the sub-radiant state ($|S\rangle$) are fit to master equation simulations, following \cref{eq:master-equation} with slight modifications. First, the qubit-drive detuning is set to zero ($\delta_i = 0$) in \cref{eq:qubit-basis-hamiltonian-sup}. Next, internal loss and dephasing terms are added to the Liouvillian, given below.

\begin{align}
\mathcal{L}_\mathrm{int}\rho = \sum_{i = 1,2} \Gamma_{\mathrm{int},i} \mathcal{D}[\hat{\sigma}_{i}]\rho
\end{align}

\begin{align}
\mathcal{L}_\mathrm{\phi}\rho = \sum_{i,j = 1,2} \frac{\Gamma_{\phi,ij}}{2}\mathcal{D}[\hat{\sigma}_{z,i},\hat{\sigma}_{z,j}]\rho
\end{align}

Here, $\Gamma_{\phi,ii}$ is the dephasing of qubit $i$ and $\Gamma_{\phi,ij} = \Gamma_{\phi,ji} = \Gamma_{\phi,\mathrm{corr}}$ is the correlated dephasing. Correlations between qubit dephasing can arise when multiple qubits are coupled to a single noise source, such as a global magnetic field \cite{Mirhosseini2019May}. The total simulated Liouvillian accounts for waveguide decay (\cref{eq:Liouvillian-full}), internal losses, and dephasing ($\mathcal{L}_\mathrm{tot}\rho = \mathcal{L}\rho + \mathcal{L_\mathrm{int}}\rho + \mathcal{L}_\mathrm{\phi}\rho$). The power spectrum of the output radiation field is then given by the two-time correlation function

\begin{align}
    S(\omega) = \text{Re} \int_{0}^{\infty}  \dfrac{d\tau}{\pi} e^{i\omega t} \langle c_\mathrm{R}^\dagger(t) c_\text{R}(t + \tau)\rangle
    \label{eq:weiner-khinchin}
\end{align}

where $c_{\mathrm{R}}$ is the collapse operator ($c_{\mathrm{R}} = \hat{\sigma}_1 + e^{ikd}\hat{\sigma}_2$) as defined in \cref{appendix:theory}. To reduce the number of free parameters in fitted simulations, we set $\Gamma_{\mathrm{int},1} = \Gamma_{\mathrm{int},2}$ and $\Gamma_{\phi,11} = \Gamma_{\phi,22}$. Fitted parameters are $\Gamma_\mathrm{int} = 0, \Gamma_{\phi} = 174 \pm 24$ kHz, $\Gamma_{\phi,\mathrm{corr}} = 127 \pm 85 $ kHz, as quoted in the main text and shown in the bottom sub-panel of \cref{figure 2}c. We note that, following the treatment given in \cite{Mirhosseini2019May}, the sub-radiant state's decay lifetime ($\mathrm{T}_1$) may be approximated as $\mathrm{T}_1 = 1/(\Gamma_\mathrm{int}+\Gamma_\phi-\Gamma_{\phi,\mathrm{corr}})$ in the case of large Purcell factor. This relation yields an estimate for $\mathrm{T}_1 = 3.4$ $\mu$s, which greatly exceeds single qubit lifetimes of $\approx 16$ ns (see \cref{appendix:lifetime}).

\section{Lifetime measurement of an individual qubit and the sub-radiant state}
\label{appendix:lifetime}
Individual qubit and resonant sub-radiant state lifetimes are presented in \cref{figure 2}d. To measure single qubit lifetimes, individual qubits are excited via the waveguide with a constant $\pi$ pulse. Qubit emission is demodulated (see \cref{appendix:tomography}) using a 80 ns time windows after a variable wait time $\tau$, and $\exc{\hat{\sigma}^\dagger\hat{\sigma}}$ is calculated. Individual qubit $\mathrm{T}_1$ lifetimes are $\mathrm{T}_1 = 16.6 \pm 0.6$ ns for qubit 1 and $\mathrm{T}_1 = 16.0 \pm 1.9$ ns for qubit 2. 

Sub-radiant state lifetimes are measured by exciting the singlet state ($|S\rangle = |eg\rangle - |ge\rangle$) with a constant $\pi$ pulse. For inter-qubit separation of $\lambda$ ($\ell = \lambda$), driving through the waveguide can only excite the super-radiant triplet state ($|T\rangle = |eg\rangle + |ge\rangle$). To excite the singlet state, XY lines are used to drive qubits out of phase simultaneously. After a variable wait time $\tau$, the ground state ($|gg\rangle$) population is then measured by state-dependent scattering \cite{zanner2022}. In this scheme, if the ground state is populated, photons are scattered between $|gg\rangle$ and $|T\rangle$, reducing the transmission amplitude. Scattering results in unit transmission if the system is fully excited to $|S\rangle$. Fits to exponential decay profiles are used to extract lifetimes. The extracted $\mathrm{T}_1$ lifetime for the singlet state is $\mathrm{T}_1 = 910 \pm 47 $ ns. We note here the discrepancy between estimated $\mathrm{T}_1$ lifetime from inelastic scattering measurements ($\mathrm{T}_1 = 3.4$ $\mu$s). We attribute this discrepancy to frequency shifts of the individual qubits over multiple days. Measured lifetimes of the sub-radiant state are crucially dependent on the individual qubit frequencies due to the interference required to protect the state from decay; frequency jitter can therefore strongly affect lifetimes. Despite these discrepancies, measured and estimated lifetimes of the sub-radiant state greatly exceed those of individual qubits, which are limited to $\approx 16$ ns.

\section{State Tomography}
\label{appendix:tomography-all}

\subsection{Measurement of qubit moments}
\label{appendix:tomography}
Most superconducting qubits use a dedicated resonator for readout using single shot dispersive readout \cite{Mallet2009}. Due to the nature of our experiment, however, the qubits are strongly coupled to a separate waveguide. Hence, without the use of a tunable coupler between the qubits and the waveguide, the qubit would mostly be decaying into this common waveguide, limiting the fidelities of the readout that we can achieve. Additionally, unless carefully designed, a readout resonator can act as another source of dissipation for the qubits, thus inhibiting the formation of a truly dark state. We instead use the strong coupling of the qubits to the common waveguide to perform our readout using field emission tomography using techniques similar to those presented in \cite{kannan_thesis}. As shown there, a direct mapping exists between the field emitted by a qubit and its state. This mapping is used to perform quantum state tomography on the two qubits. 
Following \cite{kannan_thesis}, the input-output relation for two qubits coupled to the same waveguide is given by:

\begin{align}\label{eq:out_field}
    & \hat{a}_{\mathrm{out}}(t) = \hat{a}_{\mathrm{in}}(t) \\ \nonumber
    & + e^{-i\omega_{\mathrm{1}}t_{\mathrm{1}}}\sqrt{\frac{\Gamma_{\mathrm{1D,1}}}{2}}\hat{\sigma}_{\mathrm{1}}(t) 
     + e^{-i\omega_{\mathrm{2}}t_{\mathrm{2}}}\sqrt{\frac{\Gamma_{\mathrm{1D,2}}}{2}}\hat{\sigma}_{\mathrm{2}}(t) 
\end{align}

where $\hat{\sigma}_{\mathrm{1,2}}$ are the qubit annihilation operators, $\Gamma_{\mathrm{1D,1,2}}$ are the decay rates of the respective qubits to the waveguide, and $t_{\mathrm{1,2}}$ represent the time-of-flight to the qubits. $\hat{a}_{\mathrm{out}}$ and $\hat{a}_{\mathrm{in}}$ represent the output and input propagating modes in the waveguide (this is for only one propogation direction; a similar equation exists for the opposite direction). Clearly, measuring the output field of the waveguide can also be used to gain information of the qubits' state. In fact, it can be used to determine the exact state of the qubits just before they begin emitting into the waveguide. 

To see this, consider the case of a single qubit at a frequency $\omega_{\mathrm{q}}$ whose time-dependent annihilation operator is given by $\hat{\sigma}(t)$. The output field in this case, in the absence of any drive, is given by $\hat{a}_{\mathrm{out}}(t) = \sqrt{{\Gamma_{\mathrm{1D}}}/{2}}\hat{\sigma}(t) + \hat{a}_{\mathrm{in}}(t)$, where $\hat{a}_{\mathrm{in}}(t)$ is the vacuum. For a qubit naturally decaying into a waveguide, $\hat{\sigma}(t)$ evolves as $e^{-\Gamma_{\mathrm{1D}} t/2}\hat{\sigma}(0)e^{-i\omega_{\mathrm{q}}t}$, where $\Gamma_\mathrm{1D}$ is the coupling to the waveguide. Now consider the integral, 
\begin{equation}\label{eq:cont_mode_match}
    \hat{\sigma} = \int dtf(t)\hat{a}_{\mathrm{out}}(t)e^{i\omega_{\mathrm{q}}t}
\end{equation}
where $f(t)$ represents a weighting function to give a higher weight to times where $\hat{a}_{\mathrm{out}}(t)$ is larger. This is known as temporal mode matching, and \cref{eq:cont_mode_match} is a continuous version of what actually happens during demodulation of the output field during the experiment (see \cref{appendix:mode_matching}). By choosing $f(t) = \sqrt{2\Gamma_{\mathrm{1D}}}e^{-\frac{\Gamma_{\mathrm{1D}}t}{2}}\Theta(t)$, where $\Theta(t)$ is the Heaviside step function, one can show that the integral $\hat{\sigma}$ reduces to simply $\hat{\sigma}(0)$, thus recovering the state of the qubit just before the emission began (our choice of $f(t)$ contains an additional $\sqrt{2}$ factor as compared to previous work because qubits are side-coupled rather than end-coupled to the waveguide) \cite{eichler_thesis}. Mode matching is done to maximize the detection efficiency. Thus, any choice of $f(t)$ still retains all the statistical properties of $\hat{\sigma}$ and can be used for the measurement at the cost of possibly decreased SNR. We henceforth use $\hat{\sigma}$ to denote the incoming mode, but it is equivalent to the state of the qubit at the start of the emission. 

Prior to being detected, the field $\hat{a}_{\mathrm{out}}(t)$ passes through a linear amplification chain with a net gain of $G$, usually on the order of many 10's of dB. This is because the actual detection is done using analog-to-digital converters that have a minimum voltage threshold, which is generally much higher than that of a single microwave photon. Because of the amplification, there is some inevitable noise added as well \cite{experimental_state_tomography, vinicius_thesis}, and hence what is actually detected is given by (in the limit of G $\gg$ 1),  $\hat{S} = \sqrt{G}(\hat{\sigma} + \hat{h}^{\dag})$, with $\hat{h}$ representing the added noise. We discuss how we calculate $G$ in \cref{sec:gain_calibration}. For now, consider the measurement of $\hat{S} = \hat{\sigma} + \hat{h}^{\dag}$. We are primarily interested in finding the moments of $\hat{\sigma}$, and hence need to properly account for the noise $\hat{h}$. This can be done by considering the binomial expansion of any moment of $\hat{S}$:
\begin{equation}
    \langle \hat{S}^{\dag n} \hat{S}^{m}\rangle = \sum^{n}_{i=0}\sum^{m}_{j=0}(^{n}_{i})(^{m}_{j})\langle \hat{\sigma}^{\dag i}\hat{\sigma}^{j}\rangle \langle \hat{h}^{n-i}\hat{h}^{\dag m-j}\rangle
    \label{eq:moment-formula}
\end{equation}
Here, the underlying assumption is that the noise is uncorrelated with the signal. This equation can then be used to generate a system of equations that can be used to recover the moments of $\hat{\sigma}$ instead, given that we know the moments of $\hat{h}$. To calculate the latter, we need to perform a measurement $\hat{S}_{\mathrm{0}}$ with the signal mode $\hat{\sigma}$ in vacuum, such that all the moments of $\hat{\sigma}$ are 0, except for the 0th order. This gives:
\begin{equation}
    \langle \hat{h}^{n}\hat{h}^{\dag m} \rangle = \langle \hat{S}_{\mathrm{0}}^{\dag n}\hat{S}_{\mathrm{0}}^{m} \rangle
\end{equation}
which can then be used to calculate the required moments of $\hat{\sigma}$.

Our detection protocol involves driving the two qubits for a certain duration through the waveguide, turning off the drive, and measuring the fields emitted by the two qubits. Since the two qubits are not very distant, the emitted fields are completely overlapped in the temporal domain. Instead, we note that the qubits are spectrally separated for our experimental settings and hence  $\hat{\sigma}_{\mathrm{1}}(t)$ and $\hat{\sigma}_{\mathrm{2}}(t)$ rotate at different frequencies in \cref{eq:out_field}. This allows us to use a frequency multiplexed version of the above protocol. 

Experimentally, a heterodyne detection in the microwave domain involves using an IQ mixer to down-convert from RF to a frequency range that can be digitized by an ADC. This sampled version of the field can then be demodulated (along with weights for the temporal mode matching) at the appropriate frequency to obtain an $I,Q$ pair. This $I,Q$ pair represents the two orthogonal quadratures of the detected field $S$, such that $S = I + iQ$, at the demodulated frequency. But once the signal has been downconverted and digitized, we can demodulate the signal at two different frequencies simultaneously to obtain two $I,Q$ pairs representing $\hat{S}_{\mathrm{1}}$ and $\hat{S}_{\mathrm{2}}$. These are then related to the two qubits $\hat{\sigma}_{\mathrm{1}}$ and $\hat{\sigma}_{\mathrm{2}}$ and two noise modes $\hat{h}_{\mathrm{1}}$ and $\hat{h}_{\mathrm{2}}$ in the same way that $S$ is related to $\hat{\sigma}$ and $\hat{h}$, i.e.
\begin{equation}
\begin{aligned}
    &\hat{S}_{\mathrm{1}} = \hat{\sigma}_{\mathrm{1}} + \hat{h}_{\mathrm{1}}^{\dag}, \\
    &\hat{S}_{\mathrm{2}} = \hat{\sigma}_{\mathrm{2}} + \hat{h}_{\mathrm{2}}^{\dag}.
\end{aligned}
\label{eq:noise-mode}
\end{equation}
The moments of the individual modes $\hat{\sigma}_{\mathrm{1}}$ and $\hat{\sigma}_{\mathrm{2}}$ can then be found in the same way as for $\hat{\sigma}$. The measurements for $\hat{h}_{\mathrm{1}}$ and $\hat{h}_{2}$ are done by measuring $\hat{S}_{\mathrm{01}}$ and $\hat{S}_{\mathrm{02}}$, where there is no signal mode. The cross moments such as $\hat{\sigma}_{\mathrm{1}}^{\dag}\hat{\sigma}_{\mathrm{2}}$ can also be found in a by finding the cross-correlations between $\hat{S}_{\mathrm{1}}$ and $\hat{S}_{\mathrm{2}}$. For example:
\begin{equation*}
\begin{aligned}
    \langle \hat{S}_{\mathrm{1}}^{\dag}\hat{S}_{\mathrm{2}} \rangle & = \langle (\hat{\sigma}_{\mathrm{1}}^{\dag} + \hat{h}_{\mathrm{1}})(\hat{\sigma}_{\mathrm{2}} + \hat{h}_{\mathrm{2}}^{\dag}) \rangle \\
     &= \langle \hat{\sigma}_{\mathrm{1}}^{\dag}\hat{\sigma}_{\mathrm{2}}\rangle + \langle \hat{h}_{\mathrm{1}}\hat{h}_{\mathrm{2}}^{\dag}\rangle + \langle \hat{\sigma}_{\mathrm{1}}^{\dag}\hat{h}_{\mathrm{2}}^{\dag}\rangle + \langle \hat{h}_{\mathrm{1}}\hat{\sigma}_{\mathrm{2}}\rangle \\
     \implies \langle \hat{\sigma}_{\mathrm{1}}^{\dag}\hat{\sigma}_{\mathrm{2}}\rangle &= \langle \hat{S}_{\mathrm{1}}^{\dag}\hat{S}_{\mathrm{2}} \rangle - \langle \hat{h}_{\mathrm{1}}\hat{h}_{\mathrm{2}}^{\dag}\rangle - \langle \hat{\sigma}_{\mathrm{1}}^{\dag}\rangle\langle \hat{h}_{\mathrm{2}}^{\dag}\rangle - \langle \hat{h}_{\mathrm{1}}\rangle\langle \hat{\sigma}_{\mathrm{2}}\rangle \\
\end{aligned}
\end{equation*}
again under the assumption that the noise mode $\hat{h}_{\mathrm{1,2}}$ are uncorrelated with the signal modes $\hat{\sigma}_{\mathrm{1,2}}$.

\begin{figure}[htbp]
\centering
\includegraphics[width=1\linewidth]{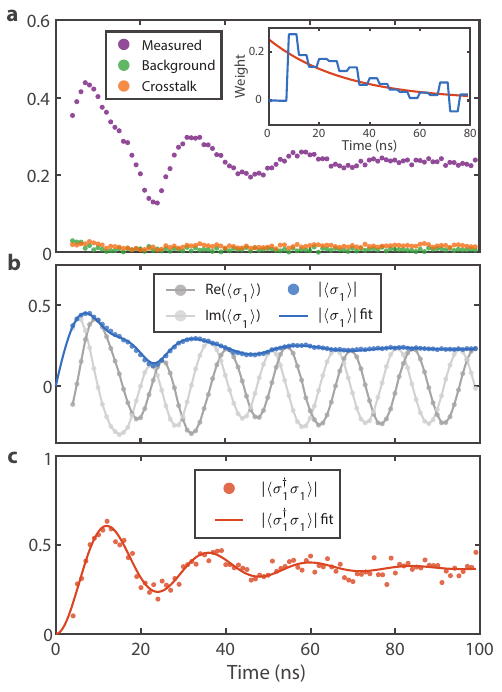}
\caption{\textbf{State tomography of an individual qubit} a) Measured field quadrature of Qubit 1 at the same detuned setting as \cref{figure 2}a, with Qubit 2 flux-tuned away (purple, $S_1$). Measured background field with both qubits flux-tuned away (green, background, $S_0$). Measured field at the same detuned setting as \cref{figure 2}a with Qubit 1 flux-tuned away (orange, crosstalk, $S_2$). The background and crosstalk are minimal compared to the signal. Inset: The mode-matched envelope of Qubit 1 (blue) and ideal exponential decay envelope (red). The envelope maximizes the signal from the qubit while rejecting noise and crosstalk sources. Background subtracted (b,c) field (photon number) measurement of Qubit 1 using mode-matched filtering, under a detuned drive (same setting as \cref{figure 2}a. Fits to master equation simulations yield Gain $= 6.3 \pm 0.1 \times 10^5$.}
\label{fig:S2_tomography}
\end{figure}

\subsection{Mode-matched filtering}
\label{appendix:mode_matching}

Our field tomography protocol recovers the qubit state by heterodyne detection of the field emitted into the waveguide. A finite time window (80 ns) is sufficient to capture the qubit emission ($T_1 \approx 16$ ns), and the emitted signal is subsequently down-converted with an IQ mixer to an intermediate frequency (IF) and digitized. The digitized signal ($\tilde{I}+i\tilde{Q}$) is then demodulated to obtain the field; the digital demodulation is detailed in the expression below.

\begin{align}
    S = I + iQ = \sum_{n} f[n](\tilde{I}[n] + i\tilde{Q}[n])e^{-i\omega_\mathrm{IF}n}
\end{align}

\noindent Here, $S$ corresponds to the measured field, $n$ is the index, and $\omega_\mathrm{IF}$ is the IF. $f[n]$ is the mode-matched filter function (discussed previously as $f(t)$ in \cref{appendix:tomography} and \cref{eq:cont_mode_match}) and is the focus of the following discussion.

In our experiments, qubits are driven via the waveguide; drives are turned off just prior to the start of the measurement window. We note that in this discussion, the qubit and drive tone are detuned at the same settings as the dark state experiment (see \cref{figure 3}a and \cref{tab:experimental_params} for details). However, we find that artifacts of the drive are present in the measurement window - mainly due to drive roll-off and drive reflections. First, drive roll-off leaks into the measurement window because of dispersion in the measurement lines. Second, repeated reflections of the drive at microwave connections in the measurement chain create delayed copies of drive pulses. In our experiment, we measure roll-off and reflections in the qubit emission window that are $\approx$ 20 dB lower than applied waveguide drives. Parasitic artifacts from the drive are within $\approx$ 5 dB of the power emitted from qubits and overlap temporally with the emission of interest. This effect therefore cannot be ignored; it is further exacerbated by the large qubit dissipation rate into the waveguide and concomitant short emission duration ($\Gamma_{\mathrm{1D}}/2\pi \approx 10$ MHz, $T_1 \approx 16$ ns). The majority of qubit emission is contained in the first 16 ns of the measurement window, which is most susceptible to the discussed parasitic effects. Capturing this drive signal in addition to qubit emission can cause un-wanted correlations between the measured and background signals, $S$ and $S_0$, compromising the measurement fidelity (see \cref{eq:moment-formula}, \cref{eq:noise-mode}). 

To overcome this problem, we optimize the mode-matching function $f[n]$ to maximize the measured qubit emission while rejecting the parasitic drive signal in the measurement window. In other words, $f[n]$ is optimized to be orthogonal to parasitic artifacts of the drive. For this purpose, we measure 10 averaged samples of single qubit emission following a 100 ns drive pulse (denoted as $S_1$). Each sample contains $10^5$ shots. Multiple averaged samples are used to mitigate effects of fluctuations in gain or qubit emission over time. We repeat this measurement with the qubit detuned away, such that the measurement window only captures parasitic drive (denoted as $S_0$) and no qubit signal. We then use non-linear least squares optimization to minimize the following loss function over the 10 samples and remove parasitic drive from the qubit signal.

\begin{align}
L_\mathrm{bkg}(f[n]) = \frac{|S_0|}{|S_1 - S_0|} 
\end{align}

Here, minimizing $|S_0|$ in the numerator orthogonalizes the filter function $f[n]$ with respect to the parasitic drive. However, because of the substantial temporal overlap between the parasitic drive and qubit signal, simply minimizing $|S_0|$ significantly reduces the qubit signal. Therefore, $|S_1 - S_0|$ is included in the denominator to maximize the output qubit signal. \cref{fig:S2_tomography}a shows the result of this optimization for qubit 1. The purple plot denotes $S_1$ and the green plot denotes $S_0$. When mode-matching is applied, we see that $S_0$ is nearly zero over a range of drive pulse durations. The optimized mode-matched temporal envelope for qubit 1 is shown in the \cref{fig:S2_tomography}a inset. We note that the filter function resembles an exponential decay with reduced weights in the first 8 ns of the measurement window. This occurs because parasitic drive artifacts are localized in this short time window. 

The steps detailed are sufficient to remove parasitic drive artifacts from the measurement window. Another source of noise in the qubit measurement arises when both qubits are measured simultaneously, as in our two-qubit dark state tomography. \cref{figure 3}a shows the detuning profile of qubit 1 and qubit 2 in the stabilization experiment. While the two qubits are separated by $\approx 3\Gamma_\mathrm{1D}$ in frequency ($|\omega_\mathrm{IF,1} - \omega_\mathrm{IF,2}| > 3\Gamma_\mathrm{1D}$), there is still the possibility for qubit demodulation to capture a ``crosstalk" signal from the adjacent qubit. To remove this crosstalk, we simultaneously optimize for a second loss function ($L_\mathrm{total} = L_\mathrm{bkg} + L_\mathrm{cross}$).

\begin{align}
L_\mathrm{cross}(f[n]) = \frac{|S_2 - S_0|}{|S_1 - S_0|} 
\end{align}

Here, the $S_2$ signal denotes the adjacent qubit. To obtain $S_2$, 10 qubit emission samples are taken with the desired qubit detuned away and the adjacent qubit at the experimental setting. $S_1$ and $S_0$ are un-changed. Similar to the previous case, minimizing $|S_2 - S_0|$ (the adjacent qubit) rejects the parasitic crosstalk from the adjacent qubit. Maximizing $|S_1 - S_0|$ prevents reduction of the desired qubit's signal. \cref{fig:S2_tomography}a shows the minimized crosstalk signal from qubit 2 (in orange), where qubit 1 is the qubit of interest. 

In the case of perfect mode matching, the mode-matching efficiency is $\eta_\mathrm{F} = 1$. For imperfect mode matching, $\eta_\mathrm{F} < 1$; this has the effect of reducing the total detection efficiency and can be interpreted in a similar way to attenuation between the device and the first amplifier \cite{eichler_thesis}. Ideally, qubit emission follows an exponentially decaying envelope, and a corresponding exponential filter function provides perfect mode matching. In our experiment, the optimized filter functions are non-exponential in order to reject the parasitic noise sources discussed above. Our experimental mode-matching efficiency to qubit emission is therefore imperfect; we find $\eta_\mathrm{F} = 0.59$ $(0.50)$ for qubit 1 (2). The \cref{fig:S2_tomography}a inset shows both the optimized (qubit 1) filter function (blue) and the corresponding ideal filter function (red). Mode-matching efficiency is calculated by taking the squared inner product between normalized ideal and optimized filter functions. We note that using an ideal exponential filter function (where $\eta_\mathrm{F} = 1$), we obtain ratios between qubit and noise signal $|S_1 - S_0|/|S_0| \approx 1$. This indicates the non-trivial effect of parasitic drive features in the absence of mode-matching optimization. On the other hand, the optimized filter function yields $|S_1 - S_0|/|S_0| \approx 20$ (see \cref{fig:S2_tomography}a).

Lastly, we verify that our mode-matching scheme does not affect the statistics of measured moments by calibrating against master equation simulations of single qubits, which is discussed in the following section.

\subsection{Gain calibration and noise temperature}
\label{sec:gain_calibration}
Qubit emission from the device passes through an amplifier chain and is then downconverted to an intermediate frequency, digitized, and digitally demodulated. The measured demodulated signal must be scaled properly to accurately recover the qubit state. We calibrate the gain of the measurement chain $G$ by fitting master equation simulations of a single qubit to demodulated emission, shown in \cref{fig:S2_tomography}b,c for qubit 1. For this measurement, qubit 1 is flux-tuned to the dark-state experiment setting (6.409 GHz) and a detuned drive (6.392 GHz) is applied. Digital demodulation is applied with $\omega_\mathrm{IF}$ corresponding to the qubit frequency. The obtained $|\exc{\hat{\sigma}_1}|_\mathrm{meas}$ and $|\exc{\hat{\sigma}_1^\dagger\hat{\sigma}_1}|_\mathrm{meas}$ are then fit simultaneously to a master equation simulation to obtain $|\exc{\hat{\sigma}_1}| = \sqrt{G}|\exc{\hat{\sigma}_1}|_\mathrm{meas}$ and $|\exc{\hat{\sigma}_1^\dagger\hat{\sigma}_1}| = G|\exc{\hat{\sigma}_1^\dagger\hat{\sigma}_1}|_\mathrm{meas}$. Gain calibration for qubit 1 and 2 yield $G = 6.3 \pm 0.1 \times 10^5$ and $G = 4.2 \pm 0.1 \times 10^5$, respectively. Note that gain values include the contribution from mode-matching. Asymmetry between calibrated gain values is attributed to asymmetric crosstalk caused by the Fano lineshape of each qubit. A qubit experiencing crosstalk signal from the neighboring qubit will have a reduced gain $G$ because mode-matched filtering will less efficiently capture the desired qubit's signal. 

We also independently extract the noise photon number and temperature of the HEMT amplifier located at the 4-K stage (see \cref{appendix:Methods}) by referencing to a resonance fluorescence measurement of a single qubit. This measurement yields a noise photon number of $n_\mathrm{HEMT} = 13$ photons and temperature of $T_\mathrm{HEMT} = 4.1$ K. Based on calculated mode-matching efficiencies, the effective noise photon numbers of our measurement are 22 (26) photons for qubit 1 (2). 

\subsection{Maximum Likelihood Estimation}
\label{appendix:MLE}
Once we have all the moments of the photonic modes emitted by the qubit (see \cref{appendix:tomography}), we use the standard method of maximum likelihood estimation (MLE) to get the density matrices for the 2-qubit state \cite{vinicius_thesis, eichler_thesis}. Since we only have emission from the first excited state of the transmon (second excited level is $\sim$300 MHz away), we  can work in the single excitation sector  for each mode. We can thus restrict our attention to the 16 moments of the form: $ A_j = (\hat{\sigma}_{1}^{\dag})^{n_1}\hat{\sigma}_{1}^{m_1}(\hat{\sigma}_{2}^{\dag})^{n_2}\hat{\sigma}_{2}^{m_2}$ $ \forall$ $ n_1, n_2, m_1, m_2 \in \{0,1\}$.

Following \cite{vinicius_thesis}, given a state $\rho$, the probability of measuring $\langle \bar{A_{j}}\rangle$ as the mean of N measurements of a particular moment $A_{j}$ is given by:
\begin{equation}
p(\langle \bar{A_{j}}\rangle|\rho) \propto e^{-|\langle \bar{A}_j\rangle - \mathrm{Tr}(A_{j}\rho)|^{2}/(v_{j}/N)}
\end{equation}
where $v_{j}$ is the variance of the moment. For large enough N, the law of large numbers tells us that we can approximate $v_{j}$ by the measured variance. One can then define the log of a likelihood function:
\begin{equation}
-\log \mathcal{L}(D|\rho) = \sum^{16}_{j=1}|\langle \bar{A_{j}}\rangle - \mathrm{Tr}(A_{j}\rho)|^{2}/(v_{j}/N)
\end{equation}
where $j$ enumerates all the different moments. This, along with the constraints that Tr($\rho$) = 1 and $\rho>0$, defines an objective function to be used for the minimization problem that finds the most likely density matrix for the given set of measured moments. 

To find the fidelity and concurrence bounds, we perform MLE on resampled copies of the moments. We record the average of all the moments over 1 million shots, and perform this entire measurement 3000 times (a total of 3 billion averages). This gives us an approximately Gaussian histogram of 3000 values, from which we extract the variance and mean. We then resample each moment from a Gaussian with the corresponding parameters to reconstruct 1000 resampled copies. For each resampled copy, we perform MLE and calculate the fidelity and concurrence. The standard deviation can be calculated from this list of fidelities and concurrences, and the 95\% confidence interval is then bounded by the 25th and 975th element of the sorted lists.

\begin{figure*}[htbp]
\centering
\includegraphics[width=0.8\linewidth]{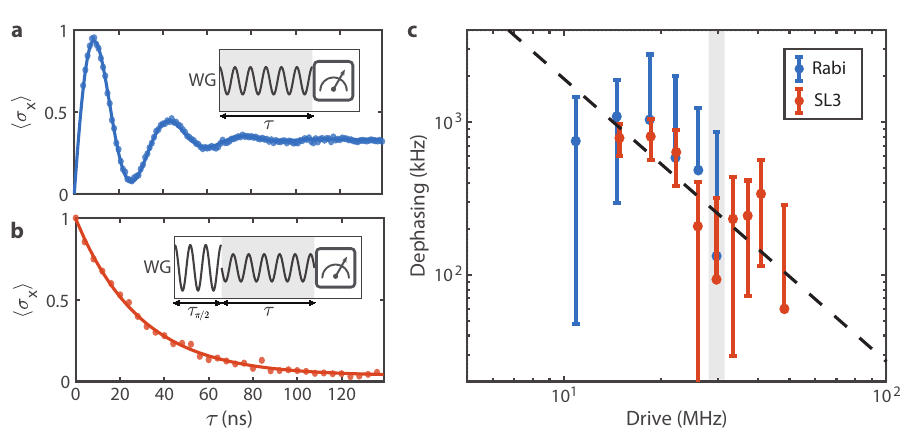}
\caption{\textbf{Individual qubit decoherence when driven via the waveguide} a) Driven Rabi oscillations of qubit 2 with drive of $\Omega_\mathrm{R}/2\pi = $ 30 MHz. Extracted $\Gamma_\phi = 133$ kHz. Inset: Rabi spectroscopy pulse sequence. b) Spin-locking protocol decay signal for qubit 2, with drive of $\Omega_\mathrm{R}/2\pi = $ 30 MHz. Extracted $\Gamma_\phi = 93$ kHz. Inset: Spin-locking protocol pulse sequence c) Extracted dephasing $\Gamma_{\nu}$ over a range of drive powers for the Rabi (blue) and SL (red) protocols, with 95$\%$ confidence intervals indicated by error bars. Dotted line indicates fit of spin-locking measurement dephasing v. drive power to $1/$f$^\alpha$, yielding $\alpha$ = 1.85. Data corresponding to a,b) are shaded in gray.}

\label{fig:S3noise}
\end{figure*}
\section{Analysis of the decoherence sources}
\label{appendix:noise}

We observe in \cref{figure 4}b that concurrences of the stabilized entangled state exceed the values predicted by qubit Purcell factors ($\sim$10, extracted from fits to transmission traces). To investigate potential sources of this discrepancy, we perform measurements of qubit decoherence caused by interaction with the waveguide bath. In particular, we study effects of the drive power.

Our dark state stabilization protocol involves applying strong drives ($\Omega/\Gamma_{\mathrm{1D}} > 1$) to the qubits via the waveguide. Previous studies have found that decoherence during a qubit's driven evolution depends on the noise power spectral density (PSD) at the Rabi frequency, which has been shown to exhibit a $1/f^\alpha$ dependence \cite{Bylander2011Jul, Yan2013Aug, Yoshihara2014Jan, Slichter2012Oct, Ithier2005Oct}. We investigate the noise PSD by measuring the driven evolution of single qubits using Rabi noise spectroscopy \cite{Bylander2011Jul} and a spin-locking (SL) protocol \cite{Yan2013Aug}.

\subsection{Driven evolution of a qubit}

The decoherence of a qubit under driven evolution has noise contributions from the qubit frequency $\omega_\mathrm{q}$, the Rabi drive frequency $\Omega_\mathrm{R}$, and quasi-static noise \cite{Bylander2011Jul}. We consider the evolution of a single qubit under a resonant drive, which is described by the following Hamiltonian in the laser frame, with $\delta = \omega_{\mathrm{q}} - \omega_{\mathrm{d}} = 0$.

\begin{align}
    \hat{H} = \frac{\delta}{2}\hat{\sigma}_z + \frac{\Omega_\mathrm{R}}{2}\hat{\sigma}_y
    \label{eq:single-qubit-hamiltonian}
\end{align}

\noindent To understand our noise spectroscopy measurement protocols, We invoke an analogy between the free and driven evolution of a qubit  \cite{Yan2013Aug}. A freely evolving qubit revolves around the z-axis of the Bloch sphere at a frequency $\delta$ (in the laser frame). The qubit will experience a longitudinal ($\Gamma_1 = 1/\mathrm{T}_1$) and transverse decay ($\Gamma_2 = \Gamma_1/2 + \Gamma_\phi = 1/\mathrm{T}_2$, where $\Gamma_\phi$ is pure dephasing) when subjected to the environment. When the qubit is driven on resonance in the y-direction, the qubit state revolves about the y-axis with the Rabi frequency $\Omega_\mathrm{R}$. In this case, the driven dynamics can be interpreted as a freely evolving spin quantized in the y-axis direction. In analogy with the case of the non-driven qubit, longitudinal ($\tilde{\Gamma}_{1}$) and transverse ($\tilde{\Gamma}_{2}$) decay rates may be defined (given below for a resonant drive) \cite{Ithier2005Oct}.

\begin{align}
    \tilde{\Gamma}_1 = \frac{1}{2}\Gamma_1 + \Gamma_\nu
\end{align}

\begin{align}
    \tilde{\Gamma}_2 = \frac{3}{4}\Gamma_1 + \frac{1}{2}\Gamma_{\nu}
\end{align}

Here, $\Gamma_{\nu} = \pi S_{z}(\Omega_\mathrm{R})$, is the pure dephasing under driven evolution, where $S_{z}(\Omega_\mathrm{R})$ is the noise PSD at the Rabi frequency $\Omega_\mathrm{R}$. ($\Gamma_\phi = \pi S_{z}(0)$ is the pure dephasing under free evolution). By varying the qubit drive strength and measuring decoherence rates, the noise PSD $S_{z}(\Omega_\mathrm{R})$ may be extracted. We measure $\tilde{\Gamma}_2$ and $\tilde{\Gamma}_1$ in two independent experiments: driven Rabi spectroscopy and a modified spin-locking (SL) protocol (respectively). We describe these experiments in the following sections.

\subsection{Driven Rabi spectroscopy}

To measure Rabi oscillations, we directly drive the qubit via the waveguide with variable pulse durations ($\tau$) and measure the emitted field to obtain the qubit state. The measurement pulse protocol is shown in the inset of \cref{fig:S3noise}a. For a qubit initially in the ground state $\exc{\sigma_z} (t=0) = -1$ and given $\Omega_\mathrm{R} \ge |\Gamma_1 - \Gamma_2|/2$, the driven evolution may be solved analytically, yielding  

\begin{align}
\begin{split}
\exc{\sigma_x}(\tau) &= x_{\infty} - \bigg(\frac{\tilde{\Gamma}_\mathrm{2} x_\infty -\Omega_\mathrm{R}}{\nu_\mathrm{R}}\sin{(\nu_\mathrm{R} \tau)} 
\\ &  
\quad + x_\infty \cos{(\nu_\mathrm{R} \tau)}\bigg) \exp({-\tilde{\Gamma}_\mathrm{2}\tau})
\label{eq:sx}
\end{split}
\end{align}

\noindent with $\exc{\sigma_y}(\tau) = 0$. Here, $\nu_\mathrm{R} = \sqrt{\Omega_\mathrm{R}^2 - (\Gamma_1 - \Gamma_2)^2/{4}}$ is the effective Rabi oscillation frequency. The steady state value of $\exc{\sigma_x}$ is $x_\infty = {\Gamma_1 \Omega}/{(\Gamma_1 \Gamma_2 + \Omega^2)}$, and the envelope decays at $\tilde{\Gamma}_2$. An example of measured Rabi oscillations of qubit 2 is shown in \cref{fig:S3noise}a, with a fit to \cref{eq:sx}. In these fits, $\Gamma_{1}$ is set constant at 10.5 MHz (estimated from transmission traces, see \cref{tab:experimental_params}), while all other parameters are allowed to vary. We note here that we exclude quasi-static noise in our analysis \cite{Bylander2011Jul}, which manifests in non-exponential decay (a feature not observed in our experiment). Sweeping the drive power $\Omega_{\mathrm{R}}$ from 10-37 MHz gives an approximate $1/f$ dependence for $\Gamma_{\phi}$ (and therefore $S_{z}(\Omega_\mathrm{R}$)), shown in \cref{fig:S3noise}c. Extracted $\Gamma_\nu$ values range from 0.1-1.1 MHz. We note that at drive powers above 37 MHz, fitted $\Gamma_{\nu}$ values approach zero. This indicates that the driven Rabi measurement is not sensitive enough to detect $\Gamma_{\phi} < 0.1$ MHz because large $\ Gamma_1$ dominates the decay. This motivates the use of the spin-locking protocol discussed next.

\subsection{Modified spin-locking protocol}

We independently extract $\Gamma_{\nu}$ using a modified spin-locking (SL) protocol discussed in \cite{Yan2013Aug}, which measures longitudinal decay in the driven qubit frame ($\tilde{\Gamma}_1$). Previous studies have demonstrated more sensitive noise spectroscopy with the SL protocol, citing robustness against low frequency fluctuations in Rabi frequency ($\Omega_\mathrm{R}$) due to instrumentation and low-frequency qubit dephasing \cite{Yan2013Aug}. The protocol involves first using a resonant $\pi/2$ pulse in the x-direction (on the Bloch sphere) to initialize a qubit state collinear with the y-axis. This is immediately followed by a continuous $\pi/2$ phase-shifted drive in the y-direction to begin the driven evolution. Ideally, the drive and qubit state are parallel, so that the qubit state relaxes to the center of the Bloch sphere at rate $\tilde{\Gamma}_1$. This pulse sequence is shown in the \cref{fig:S3noise}b inset. The qubit state is measured via waveguide emission as it decays. We note here that our measurement protocol does not include a second $\pi/2$ pulse used in \cite{Yan2013Aug} because we do not use a readout resonator. An example of the exponential decay profile measured for qubit 2 is shown in \cref{fig:S3noise}b. Fits to the relaxation include fixed parameter $\Gamma_1 = 10.5$ MHz and $\Gamma_\nu$ free. Extracted $\Gamma_\nu$ ranges from 60-800 kHz over a range of Rabi frequencies 15-48 MHz. A fit to extracted $\Gamma_\nu$ reveals a $1/f^\alpha$ dependence with $\alpha = 1.85$.

\begin{figure}[htbp]
\centering
\includegraphics[width=\linewidth]{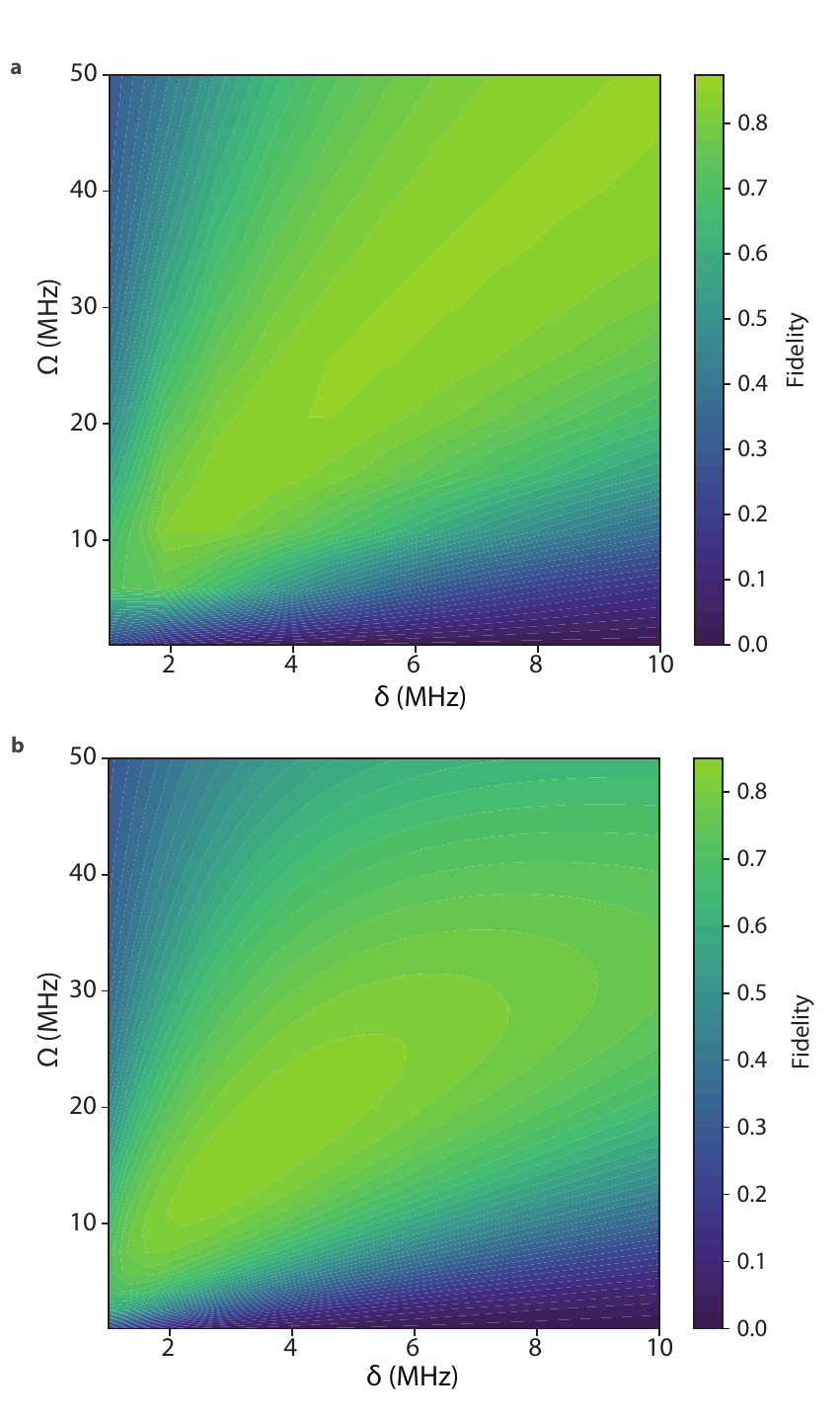}
\caption{Achievable fidelities for different parameter regimes of drive strength ($\Omega$ and detunings ($\delta$), without the e-f transition (a) and with the e-f transition (b) included in the master equation. Clearly, the inclusion of the third level makes a significant difference in the operational region, while the maximum fidelity is roughly the same. See \cref{appendix:ef} for more details of simulation.}
\label{fig:ef_comparison}
\end{figure}

\subsection{Purcell factor under driven evolution}\label{appendix:driven_purcells}

The observed reduction in $\Gamma_{\nu}$ at higher drive powers indicates that the true Purcell factor of our qubits is larger than $\sim$10 (extracted from VNA traces), which qualitatively explains the increase in concurrence observed in \cref{figure 4}b. However, extracted concurrences do not indicate monotonically increasing Purcell factors with drive power (\cref{figure 4}b), which would be expected for a purely $1/f^\alpha$ trend in dephasing. Instead, at low powers, concurrences indicate Purcell factors exceeding 30, while at high powers, Purcell factors range from 10-30. Previous studies \cite{Yan2013Aug} have reported Lorentzian ``bump-like" features in noise PSD attributed to coherent two-level system (TLS) fluctuators. While this could explain the above discrepancies, the large confidence bounds of extracted $\Gamma_\nu$ values (\cref{fig:S3noise}c) preclude a quantitative conclusion.

\section{Numerical modeling}\label{appendix:simulation}

We use a master equation solver in QuTip \cite{qutip, qutip2} to simulate our system.  We use the same Hamiltonian as in \cref{appendix:theory}, but also include the third level (f) of transmons (see \cref{appendix:ef}). The simulation also includes dissipators for dephasing and intrinsic loss of the individual qubits (for the second level), as well as for correlated decay into the waveguide between the two transmons. Note that we do not include any waveguide mediated interaction in the Hamiltonian as discussed in \cref{appendix:theory}. The shaded regions in \cref{figure 3} and \cref{figure 4} were found via these simulations, using different Purcell factors. The Purcell factor is varied by changing the intrinsic loss and dephasing rates of the individual transmons. Parameters such as the individual coupling rates of the qubits to the waveguide are set to match the fits of  experimental data at the operation frequencies, i.e. coupling rates of 8.7 MHz for Qubit 1 and 10.5 MHz for Qubit 2.

\subsection{Effects of the e-f transition}\label{appendix:ef}
An important aspect of the simulations was the inclusion of the third level of the transmons. We observed that while the third level did not significantly impact the maximum achievable fidelities, it did affect the operation ranges for a given fidelity. As can be seen from \cref{appendix:theory}, for no dephasing and no intrinsic loss, the singlet fraction, which is directly related to the fidelity for a pure state, is only a function of the ratio of the detuning and the drive power. This would mean that increasing the drive strength for any detuning would help increase the fidelity of the steady state. Practically, this would also increase the settling time which competes with any intrinsic loss or dephasing of the qubits. Hence, there are optimal drive-to-detuning ratios for a true two-level system model. This can be seen in panel (a) of \cref{fig:ef_comparison} which shows results from a simulation with a Purcell factor of 200 for the qubits and equal couplings of 10 MHz to the waveguide. We perform a sweep for various drives and detunings, but do not include the e-f transition for panel (a). Here, we can clearly see regions of high fidelity even for high drive powers. In contrast, once we include the e-f transition (panel (b)), the parameter space with high fidelities is restricted. Increasing the drive power for a given drive-to-detuning ratio does not necessarily help in this case. We expect that this has to do with the e-f transition allowing the system to escape out of the relevant Hilbert space. Since the third level is populated more for higher drive powers, we see a clear deviation from the two-level system model at these drives. 

\section{Achievable fidelities in future experiments}\label{appendix:achievable_fidelities}
In this section we discuss the challenges that we faced in our experiment that limited our fidelities. We will then discuss avenues to overcome these bottlenecks and use simulations to show that it is possible to even reach fidelities of over 90\% with this driven-dissipative stabilization protocol and experimentally viable parameters for the transmons. 
\begin{figure}[htbp]
\centering
\includegraphics[width=\linewidth]{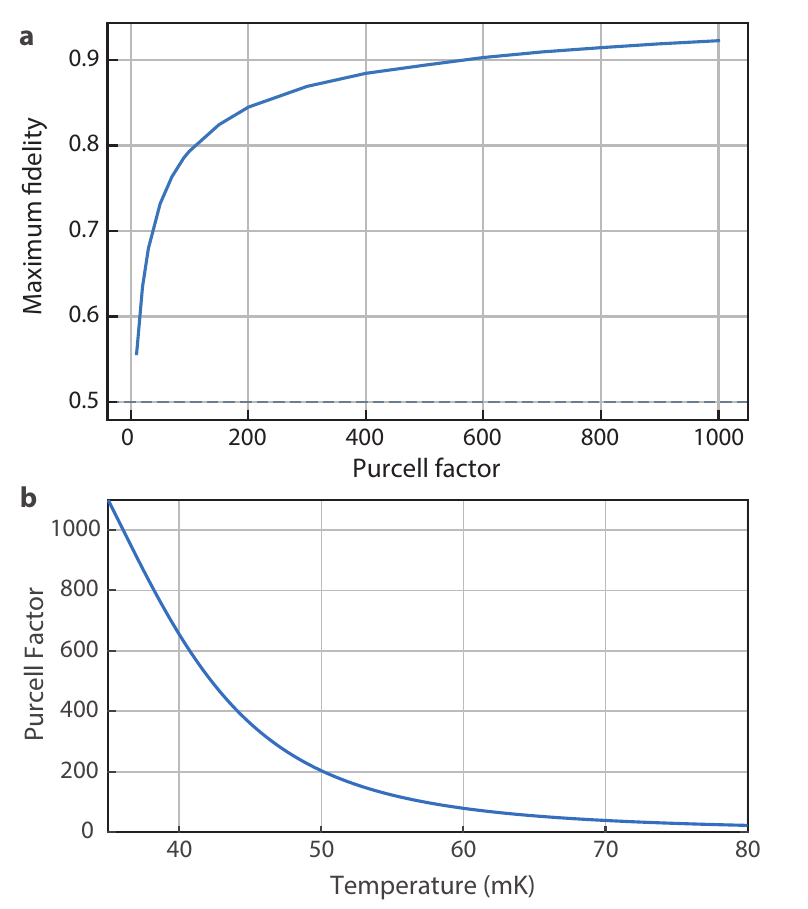}\caption{Maximum achievable fidelities a) Achievable fidelities for different Purcell factors of the individual qubits. Dashed line marks the 50\% threshold for entanglement. b) Estimates of maximum achievable Purcell factors under strong driving. Upper limit on Purcell factor set by waveguide temperature when dephasing is ignored, assuming internal losses of $\Gamma^\prime = 6$ kHz. At 39 mK, $\Gamma_\mathrm{1D}/\Gamma^\prime$ = 730.}
\label{fig:outlook}
\end{figure}

The first limitation of our experiment was in fact the effect of the e-f transition. As mentioned in \cref{appendix:ef}, the highest achievable fidelity is not very different between a true two-level system model and that of a transom, but the operation point does change and tends to move towards lower detunings. While we could in principle use such an operation point, the effects of cross-talk and drive roll-off (described in \cref{appendix:mode_matching}) were more pronounced at lower detuning and would have affected our measurements. 

This effect could be avoided by using fast flux lines to tune the qubits far enough after state preparation to conduct the measurement. Alternatively, dedicated readout resonators with other architectural changes to the design could be used as well. In particular, the coupling of the qubits to the waveguide would have to be tunable. Without a tunable coupling, the qubits would emit into the waveguide during readout which would reduce the fidelities of measurement. The fact that the qubits are strongly coupled to the waveguide (order of 10’s of MHz to increase the Purcell factors) would reduce the fidelities further since they impose stringent limitations on the duration of the dispersive readout pulse. 

Another limitation to our experiment was the imbalance in $\Gamma_\mathrm{1D}$ of the two qubits at their operation points due to standing modes in the background causing Fano in the waveguide response. While this in itself is not very detrimental to the fidelities, having a common drive and different $\Gamma_\mathrm{1D}$ causes a difference in the Rabi drives that each qubit sees. This imbalance in drive strength can be shown to prevent the singlet state from being truly dark \cite{10.1103/physreva.91.042116}, thus reducing the fidelities. Our experiment had an imbalance of around 4\%, which along with the limitation on detunings mentioned earlier, reduced our maximum achievable fidelities by around 6\% (see blue shaded region of \cref{figure 4}c). The previously mentioned design of having tunable couplers can help mitigate this effect as the coupling rates would then be in-situ tunable. 

While these two challenges did affect our fidelities, the primary bottleneck in our experiment was in fact the Purcell factor of our qubits ($P_\mathrm{1D}$ = $\Gamma_\mathrm{1D}/\Gamma', \Gamma' = 2\Gamma_2 - \Gamma_\mathrm{1D} = \Gamma_\mathrm{int} + 2\Gamma_\phi$). This sets an upper bound on the fidelities in our stabilization protocol. \cref{fig:outlook}a shows the maximum achievable fidelities as the Purcell factor is increased. From the measured transmission of our qubits, we estimate that our (undriven) Purcell factors are $~$10, which clearly limits our fidelities. Of course, in our protocol, the continuous driving of the qubits reduces the pure dephasing of the qubits (\cref{appendix:driven_purcells}) and hence we see higher concurrences and fidelities than that estimated by a Purcell factor of only 10. 

We estimate that this scheme of generating entanglement can reach fidelities exceeding 90$\%$ (concurrence exceeding 0.86) for driven Purcell factors of 600 and above. To our knowledge, the highest reported Purcell factor for a qubit directly coupled to a waveguide is approximately 200 \cite{Mirhosseini2019May}. Meanwhile, state-of-the-art aluminum transmons exhibit lifetimes of several hundred microseconds \cite{Biznarova2023Oct}, which correspond to Purcell factors exceeding 1000. One of the causes of this apparent gap in performance is the thermal occupation of the waveguide in the former case. Qubits directly coupled to waveguides experience additional decoherence due to interaction with thermal photons residing in the photonic bath. This non-zero waveguide temperature degrades the Purcell factor. Again, for our stabilization protocol, strong drives mitigate this effect slightly due to a reduction in pure dephasing (discussed in \cref{appendix:driven_purcells}). Nevertheless, the finite waveguide temperature ultimately imposes an upper bound on achievable Purcell factors. Following the master equation treatment of \cite{Mirhosseini2019May}, Purcell factor in the presence of a finite waveguide temperature is $P_\mathrm{1D} = \Gamma_\mathrm{1D}/\Gamma^\prime_\mathrm{th}$, where $\Gamma^\prime_\mathrm{th}$ is the intrinsic qubit decoherence and is defined as $\Gamma^\prime_\mathrm{th} = 2\Gamma_{2,\mathrm{th}} - \Gamma_\mathrm{1D}$ ($\Gamma_{2,\mathrm{th}} = \Gamma_{1,\mathrm{th}}/2+\Gamma_\phi$, $\Gamma_{1,\mathrm{th}} = (2\bar{n}_\mathrm{th}+1)\Gamma_1$, $\bar{n}_\mathrm{th}$ denotes bath occupancy).

Using a reference measurement of $\Gamma_1$ for a separate qubit chip (without any coupling to a waveguide), we may estimate $\Gamma^\prime$ for the main device. This separate chip contains a qubit coupled to a readout resonator and undergoes an identical fabrication procedure as the main device. We measure $T_1 = 25.5 \pm 1~ \mu$s for the separate qubit device, corresponding to $\Gamma^\prime/2\pi$ of 6 kHz. Previously reported waveguide temperatures have been as low as 39 mK \cite{kannan_thesis}. At 39 mK, the Purcell factor is limited to 730 by the bath (for an estimate of $\Gamma^\prime/2\pi = 6$ kHz), as shown in \cref{fig:outlook}b. This would enable us to reach a fidelity of 91\% and a concurrence of 0.88 with a stabilization rate of 250 kHz.

\end{document}